\documentclass[onecolumn,preprint,3p,authoryear]{elsarticle}

\usepackage{tikz,amsfonts,amssymb,amsmath,algorithm,algpseudocode,subcaption}

\begin{document}
\begin{frontmatter}
\title{Greedy Stein Variational Gradient Descent: An algorithmic approach for wave prospection problems}
\author{José L. Varona-Santana\thanks{Corresponding author}}
\ead{jose.varona@cimat.mx}
\affiliation[1]{organization={Centro de Investigación en Matemáticas A.C},
    addressline={ De Jalisco s/n, Valenciana},
    postcode={36023 MX},
    city={Guanajuato},
    country={México}}

\cortext[cor1]{José L. Varona-Santana}

\author{Marcos A. Capistr\'an}
\ead{marcos@cimat.mx}
\affiliation[1]{organization={Centro de Investigación en Matemáticas A.C},
    addressline={ Parque Científico y Tecnológico de Yucatán
    Km 5.5 Carretera Sierra Papaca},
    postcode={97302 MX},
    city={Yucatán},
    country={México}}

\begin{abstract}

    In this project, we propose a Variational Inference algorithm to approximate posterior distributions. Building on prior methods, we develop the Gradient-Steered Stein Variational Gradient Descent (G-SVGD) approach. This method introduces a novel loss function that combines a weighted gradient and the Evidence Lower Bound (ELBO) to enhance convergence speed and accuracy. The learning rate is determined through a suboptimal minimization of this loss function within a gradient descent framework.

    The G-SVGD method is compared against the standard Stein Variational Gradient Descent (SVGD) approach, employing the ADAM optimizer for learning rate adaptation, as well as the Markov Chain Monte Carlo (MCMC) method. We assess performance in two wave prospection models representing low-contrast and high-contrast subsurface scenarios. To achieve robust numerical approximations in the forward model solver, a five-point operator is employed, while the adjoint method improves accuracy in computing gradients of the log posterior.

    Our findings demonstrate that G-SVGD accelerates convergence and offers improved performance in scenarios where gradient evaluation is computationally expensive. The abstract highlights the algorithm’s applicability to wave prospection models and its potential for broader applications in Bayesian inference. Finally, we discuss the benefits and limitations of G-SVGD, emphasizing its contribution to advancing computational efficiency in uncertainty quantification.
\end{abstract}


\end{frontmatter}

\section{Introduction}\label{sec:introduction}

Our research investigates geophysical applications of wave prospecting models in one-dimensional scenarios, with a focus on determining the velocity field in low-contrast and high-contrast stratified media. To achieve this, we propose a Bayesian model that describes the parameter space density, providing insights into the velocity  field.

Sampling from the posterior distribution is central to Bayesian inference but poses significant computational challenges, especially in high-dimensional parameter spaces. Markov Chain Monte Carlo (MCMC) methods, while standard, are computationally expensive in such cases. Variational Inference (VI) offers an alternative by approximating the posterior through optimization, minimizing the Kullback-Leibler (KL) divergence between a test distribution and the posterior.

VI methods have become widely used in machine learning and artificial intelligence due to their scalability and efficiency. These methods often employ gradient descent to optimize the Evidence Lower Bound (ELBO), improving convergence speed. \citep{ruder2016overview} provides a comprehensive review of gradient descent methods, emphasizing their applicability in optimizing complex models. \citep{liu2016stein} introduced Stein Variational Gradient Descent (SVGD), a particle-based variational inference method that iteratively updates a set of particles to approximate a target posterior distribution. SVGD employs functional gradient descent to minimize the KL divergence between the target and the particle distribution, leveraging Stein's identity and kernelized Stein discrepancy. This method combines smooth transformations of particles with gradients derived from Stein's operator, ensuring convergence guarantees while enabling efficient approximation of posterior distributions through both gradient-based updates and repulsion between particles to prevent collapse. \citep{guo2016boosting} proposed a boosting variational inference (BVI) method to approximate the posterior distribution. The method starts with a single component and incrementally adds components to improve approximation accuracy. BVI employs a gradient boosting framework and uses Taylor expansion to derive functional gradients for updating components. The learning rate is determined using a stochastic Newton's method, which estimates first and second derivatives of the objective function via Monte Carlo. Gradients and Hessians of the residual log-likelihood are evaluated, with the Hessians computed through numerical approximation. The method faces computational challenges in determining the learning rate, as it requires a stochastic minimization of the KL divergence at each step. \citep{blei2017variational}  minimizes the KL divergence between a chosen family of approximate densities and the target posterior, using the ELBO as the optimization objective. Compared to MCMC, VI is faster and more scalable, particularly for large datasets and complex models. Key techniques include mean-field VI, which assumes independence among latent variables for simplification, and extensions like Coordinate Ascent Variational Inference (CAVI) and Stochastic Variational Inference (SVI), which enable efficient computations and scaling to massive data. Advanced methods such as Stein Variational Gradient Descent (SVGD) and boosting approaches enhance flexibility and performance in high-dimensional parameter spaces. \citep{locatello2018boosting} introduce a modification to the Frank-Wolfe algorithm to approximate posterior distributions by projecting onto the convex hull of a test family of densities. Using an affine invariant adaptation, the algorithm improves convergence speed by optimizing within the geometry of the convex hull. The objective is to minimize the KL divergence between the approximate and true posterior, with guarantees for sublinear and linear convergence under specific conditions.
\citep{egorov2019maxentropy}  tackles the challenge of variational inference by introducing MaxEntropy Pursuit Variational Inference, a method aimed at improving the approximation of complex multimodal posterior distributions. The approach minimizes the KL divergence under constraints defined by a test family of distributions \( Q \). Recognizing the limitations of fixed \( Q \), it incorporates Maximum Entropy regularization to guide optimization, enabling more flexible posterior approximations. The iterative refinement of the variational distribution is achieved via an affine update rule:
\begin{equation}
    q_{t+1} = (1 - \alpha) q_t + \alpha h, \quad \alpha \in (0, 1), \, h \in Q    
\end{equation}
where \( h \) is selected from the test family to maximize the  ELBO increment. This method allows for controlled complexity growth of the posterior by iteratively adding components, ensuring efficient computation.  \citep{futami2019bayesian}  addresses the challenge of approximating posterior distributions in Bayesian inference for complex models such as neural networks. It introduces a novel method called Maximum Mean Discrepancy minimization via the Frank-Wolfe algorithm (MMD-FW), which combines particle-based approximation with convex optimization to achieve computational efficiency and theoretical guarantees. \citep{brennan2020greedy} presents a novel framework for solving high-dimensional Bayesian inference problems using structure-exploiting lazy maps. These maps operate in a low-dimensional subspace, enabling efficient and tractable approximations of complex posterior distributions. The framework identifies low-dimensional structures by minimizing an upper bound on the KL divergence, leveraging logarithmic Sobolev inequalities to focus computational efforts on significant discrepancies in the posterior.

In our work, we propose a VI algorithm with theoretical guarantees of convergence. We build on the hypotheses of \citep{liu2016stein} and leverage the benefits of the ELBO as described in \citep{blei2017variational}. Our approach addresses three key gaps identified in \citep{liu2016stein}: the initial set of particles, learning rate, and stopping condition. We improve convergence to the posterior distribution by employing Kernel Density Estimation (KDE) as the test distribution.
We trade off accuracy in the posterior approximation to enhance convergence toward a Gaussian distribution centered around the local modes of the posterior. To achieve this, we introduce a loss function that combines the ELBO and the gradient of the log-posterior. Furthermore, we reduce numerical errors in gradient evaluation by employing the adjoint method, restricting the error to the numerical accuracy of the PDE solver. To further minimize this error, we use the five-point operator, which increases the accuracy of numerical derivative approximations.
Our method finds a suboptimal solution for the learning rate by minimizing the proposed loss function, ensuring efficient convergence while balancing the trade-offs between accuracy and computational efficiency.

The document is structured as follows: 

In section \ref{sec:Theoretical_Framework}, we cover the theoretical foundations and the description of our algorithm. We address the wave equation and the method to solve it. Then, we explain the five-point operator and its derivation from Taylor's polynomial. This is followed by the forward model of our investigation, differentiating between the surrogate model and the continuous model. Later, we address the weighted gradient and Bayesian Inference via Variational Gradient Descent \citep{liu2016stein}. We conclude the section with the introduction of the Loss Function, the Greedy Stein Variational Algorithm, and our representation of the Adjoint Method applied in our context.
In section 3, we discuss the numerical results. First, we compare the derivative approximation using the traditional finite difference method and the five-point operator. Then, we present the numerical results of our approach applied to two wave prospecting models.

\section{Theoretical Framework}\label{sec:Theoretical_Framework}

This section outlines the theoretical foundations supporting this project. It begins with exploring the wave equation, addressing its formulation in homogeneous and non-homogeneous media, and the principle of superposition of waves. The five-point operator is introduced, highlighting its role in the numerical approximation of first- and second-order derivatives. Then, we address the forward model for the two examples of prospection models.
The section then transitions to the Stein Variational Gradient Descent and the step size in the gradient descent method. Subsequently, the ELBO is examined as a key element in the proposed approach. The loss function is explained in detail, emphasizing its importance in optimization. Finally, the section concludes with an introduction to the G-SVGD algorithm, which integrates these theoretical concepts into a cohesive computational framework and the adjoint method applied to the context of our wave prospection models.

\subsection{Straitfied Medium}\label{sec:Straitfied_Medium}
In the rest of this document, we make use of the following definitions:

\begin{description}
    \item[\textbf{Stratified medium}:] A non-homogeneous medium composed of different materials, where the physical properties vary from one point in space to another.
    \item[\textbf{High-contrast stratified medium}:] A medium in which the physical properties of its components exhibit radical changes from one point in space to another.
    \item[\textbf{Low-contrast stratified medium}:] A medium in which the physical properties of its components exhibit smooth changes from one point in space to another.
\end{description}

These are the changes we aim to describe in our investigation. In the context of velocity, we define a high-contrast stratified layer as one where the transverse velocity field of waves traveling through it is described by a stepwise function. In such a layer, the transitions between regions cannot be represented by a smooth function, as the materials constituting the structure prevent a smooth transition of the wave as it propagates.

\begin{figure}
    \includegraphics[width=\textwidth]{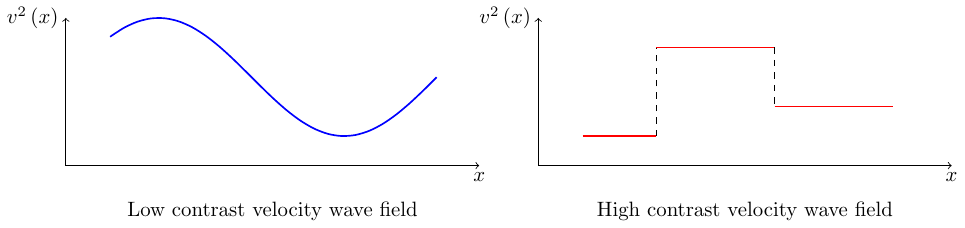}
    \label{fig:sinusoidal_step}
    \caption{Left: The velocity field is represented by a smooth curve, illustrating a low-contrast stratified medium; Right: The velocity field is represented by a stepwise function, illustrating a high-contrast stratified medium.}
\end{figure}

Such a structure can be described as a composition of other regions, where each constituent part is a low-contrast stratified layer. On this scenario, we consider the case where these low-contrast regions are homogeneous. A low-contrast stratified layer refers to regions where the transitions in the transverse velocity field of waves are described by smooth functions. Waves traveling through this type of structure exhibit smooth transitions from one spatial point to another, with controlled changes in their amplitude. The clearest example of such a layer is a homogeneous region. Figure \ref{fig:sinusoidal_step} illustrates the transverse velocity field for each scenario.

\subsection{Wave equation}
We use the wave equation to describe the velocity field of waves traveling within each of these layers. With the wave equation as our physical governing model, we construct two wave prospection models to describe the scenario of a static material perturbed by an external force. These external forces are modeled as waves entering the structure at each boundary and traveling uninterrupted out of it. By "uninterrupted," we mean that no additional forces or perturbations are present in the system other than those provided by the boundaries and the elastic tension of the materials. 

For high-contrast scenarios, we assume that each homogeneous subregion is perturbed by two waves traveling in opposite directions, and the entire structure is treated as the continuous union of these subregions. This setup allows for the exploration of interactions between areas with differing compositions. Such exploration is essential to determine the reflection and transmission of waves traveling through the medium, as well as their propagation and amplitude. Similar approaches were developed by \citep{capistran2012full,capistran2020reliability}, where the region of interest was modeled as a composition of homogeneous layers.

For low-contrast scenarios, we consider a single region perturbed by a wave entering from one boundary, while the wave travels uninterrupted out of the area of interest through the other boundary. This setup allows for the investigation of the behavior of traveling waves in a non-homogeneous region. In this case, careful consideration is required for the discrete approximation of the differential model. We adopted the \textit{Displacement} formulation of the 1D wave equation \citep{moczo1998introduction}, excluding the damping term for simplicity.

\subsubsection{One-Dimensional Wave Equation in a Homogeneous Layer}\label{sub:high_contrast_medium}

For a high-contrast stratified structure, we model a static layer perturbed by an external source, where the resulting disturbance propagates without interruption and exits through open boundaries. The displacement-based formulation for this scenario is:

\begin{equation}
    \begin{array}{rlrl}
        \partial_{t}^{2}u_{r} - c^2 \partial_{x}^{2}u_{r} & = 0,                            & x \in (0, L), & t \in (0, T] \\
        u_{r}(x, 0)                                         & = 0,                            & x \in [0, L]                \\
        \partial_{t}{u_{r}}(x, 0)                           & = 0,                            & x \in [0, L]                \\
        u_{r}(0, t)                                         & = F_{l}(t),                     & t > 0                      \\
        \partial_{t}u_{r}(L, t)                             & = -c\partial_{x}u_{r}(L, t),    & t > 0
    \end{array}
    \label{eq:DF1DWE_ur}
\end{equation}

\begin{equation}
    \begin{array}{rlrl}
        \partial_{t}^{2}u_{l} - c^2 \partial_{x}^{2}u_{l} & = 0,                            & x \in (0, L), & t \in (0, T] \\
        u_{l}(x, 0)                                         & = 0,                            & x \in [0, L]                \\
        \partial_{t}{u_{l}}(x, 0)                           & = 0,                            & x \in [0, L]                \\
        \partial_{t}u_{l}(0, t)                             & = c\partial_{x}u_{l}(0, t),     & t > 0                      \\
        u_{l}(L, t)                                         & = F_{r}(t),                     & t > 0
    \end{array}
    \label{eq:DF1DWE_ul}
\end{equation}
The equations \eqref{eq:DF1DWE_ur} and \eqref{eq:DF1DWE_ul} describe the same static structure perturbed by external forces \( F_l \) and \( F_r \). This system models two waves propagating in opposite directions within a homogeneous layer with absorbing boundaries. The first-order open boundary conditions, \( \partial_{t}u_{r}(L, t) + c\partial_{x}u_{r}(L, t) = 0 \) and \( \partial_{t}u_{l}(0, t) - c\partial_{x}u_{l}(0, t) = 0 \), are derived from the equations proposed in \citep{buckman2012onesided}. To model the wave behavior, we apply the superposition principle.

\paragraph{Wave Superposition Principle}
The principle of superposition states that when two or more waves propagate simultaneously through the same layer, the resulting displacement is the algebraic sum of the individual wave displacements \citep[section 16-5: Interference of waves]{halliday2013fundamentals}. 

If \( u_l \) and \( u_r \) are the only waves present in the layer bounded by \(\left[ 0, L \right]\), the resulting wave is:
\begin{equation}
    u(x, t) = u_r(x, t) + u_l(x, t)
    \label{eq:wave_solution_homogeneous}
\end{equation}

\subsubsection{One-Dimensional Wave Equation in Non-Homogeneous Layer}
Our modeling scenario in this case is similar to the previous section, but now we consider a non-homogeneous layer. This means that the transverse velocity of the wave depends on space, which introduces modifications to the wave equation system. The system describing a wave perturbing a non-homogeneous static medium, with one open boundary and the other acting as a source, is given by:

\begin{equation}
    \begin{array}{rlrl}
        \partial_{t}{2}{u} - \nabla \cdot\left( c^2\left( x \right)\nabla_{x}u \right) & = 0,                            & x \in (0, L), & t \in (0, T] \\
        u(x, 0)                                         & = 0,                            & x \in [0, L]                \\
        \partial_{t}{u}(x, 0)                           & = 0,                            & x \in [0, L]                \\
        \partial_{t}u(0, t)                             & = F_{l}(t),     & t > 0                      \\
        u(L, t)                                         & = -c_{L}\partial_{x}u(0, t),                     & t > 0
    \end{array}
    \label{eq:DF1DWE_non-homogeneous}
\end{equation}
The wave equation is a hyberbolic partial differential equation that its sensitive to jumps on the velocity field. Meaning that the numerical solution for this system need to be treated carefully.

\subsection{The five points finite differences operator}

The finite difference method is a numerical approach for solving differential equations by approximating the derivatives of the functions within the system. The numerical accuracy of this method, particularly when applied to wave and transport equations, has been well-documented and demonstrated by several authors \cite{quarteroni2006numerical,quarteroni2008numerical,langtangen2016finite,igel2017computational}. These studies highlight the effectiveness of the finite difference method in approximating numerical solutions to partial differential equations. This method is widely used in computational seismology, where it is often implemented as a \textit{five-point operator} to approximate the spatial derivative.

Applying this numerical approach to solve the wave equation involves two key considerations. The first is the Courant number, also known as the Courant-Friedrichs-Lewy (CFL) condition. This number is used to establish the relationship between temporal and spatial resolutions to ensure numerical stability in the system's solution. Once the CFL condition is satisfied, it is crucial to determine the spatial resolution to ensure the accuracy of the approximation.

Traditional approaches in the finite difference method include the centered difference and backward difference methods. Both are third-order approximations of the derivative of a function based on Taylor's polynomial expansion.

Let 
\[
\boldsymbol{\alpha}_{i} = \begin{pmatrix} \alpha_1^{i} \\ \alpha_2^{i} \\ \alpha_3^{i} \\ \alpha_4^{i} \\ \alpha_5^{i} \end{pmatrix}, 
\]
a column vector of real coefficients, and
\[
\boldsymbol{T}_{i} = \begin{pmatrix} T^{4}_{i-2} \\ T^{4}_{i-1} \\ T^{4}_{i} \\ T^{4}_{i+1} \\ T^{4}_{i+2} \end{pmatrix},
\]
a column vector, where each component represents the 4th-degree Taylor polynomial center at \( x_i \) and evaluated in \(x_j\).
The \( n \)-order derivative and the coefficient values for the five-point operator are obtained from the Taylor polynomial using the following relation:
\begin{equation}
\frac{\partial^n f\left(x_i\right)}{\partial x^n} = \boldsymbol{\alpha}_{i}^\top \boldsymbol{T}_i + R_i\left(\boldsymbol{\alpha}, \Delta x, f^{\left(5\right)}\right),
\end{equation}
where the error in the approximation, using the Lagrange version of the remainder, is given by:
\begin{equation}
    \left| \frac{\partial^n f\left(x_i\right)}{\partial x^n} - \boldsymbol{\alpha}^\top \boldsymbol{T}_i \right| = \left| R\left(\boldsymbol{\alpha}, \Delta x, f^{\left(5\right)}\right) \right| \leq \frac{\left| f^{\left(5\right)}\left(c\right) \right| \left( 2\Delta x \right)^{5}}{5!} \left( \left| \alpha_1 \right| + \left| \alpha_2 \right| + \left| \alpha_4 \right| + \left| \alpha_5 \right| \right)
\label{eq:error5}    
\end{equation}
where \( c \in \left[ x_{i-2},x_{i+2} \right]\) is a point within the interval of approximation.
If \(\varepsilon\) is the desired error in the approximation, then the spatial step size \(\Delta x\) must satisfy:
\[
\Delta x \leq \frac{1}{2} \sqrt[5]{\frac{120\varepsilon}{\left| f^{\left(5\right)}\left(c\right) \right| \left( \left| \alpha_1 \right| + \left| \alpha_2 \right| + \left| \alpha_4 \right| + \left| \alpha_5 \right| \right)}}.
\]

Consequently, the associated system of equations is given by:

\begin{equation}
    \mathbf{A}\left(\phi_i\right) \boldsymbol{\alpha} = \mathbf{f}^{(n)} \approx \frac{\partial^n f\left(x_i\right)}{\partial x^n} 
    \label{eq:5po_discrete_system}
\end{equation}
where the system matrix, \( \mathbf{A} \), depends on the stencil \( \phi_i \) used for the approximation, which in turn depends on the node's position within the discretized domain. The vector \( \mathbf{f}^{(n)} \) represents the \( n \)-th derivative to be approximated. The maximum order of the derivative that can be approximated is equal to the stencil size, as determined by the formulation of the system of equations using the Taylor polynomial expansion on the selected stencil.

In our work, we identified five fundamental stencils and the associated systems of equations.

\begin{enumerate}

    \item  $\phi_0=\left[x_{0},x_{1},x_{2},x_{3},x_{4}\right]$
    This corresponds to the node \( x_0 \) located at the left boundary.

\begin{equation}
        \begin{pmatrix}
            1&1&1&1&1 \\
            0&1&2&3&4 \\
            0&1&4&9&16 \\
            0&1&8&27&64 \\
            0&1&16&81&256 
        \end{pmatrix}
        \boldsymbol{\alpha}_{0}
        =
        \begin{pmatrix}
            0&0&0&0\\
            \frac{1}{dx}& 0&0&0\\
            0& \frac{2!}{dx^2}&0&0 \\
            0 &0& \frac{3!}{dx^3}&0 \\
            0 &0&0&\frac{4!}{dx^4}
        \end{pmatrix}
\end{equation}

\item  $\phi_1=\left[x_{0},x_{1},x_{2},x_{3},x_{4}\right]$
This corresponds to the node \( x_1 \).

\begin{equation}
        \begin{pmatrix}
            1   &1  &1  &1  &1 \\
            -1  &0  &1  &2  &3 \\
            1   &0  &1  &4  &9 \\
            -1  &0  &1  &8 &27\\
            1   &0  &1 &16 &81 
        \end{pmatrix}
        \boldsymbol{\alpha}_{1}
        =
        \begin{pmatrix}
            0&0&0&0\\
            \frac{1}{dx}& 0&0&0\\
            0& \frac{2!}{dx^2}&0&0 \\
            0 &0& \frac{3!}{dx^3}&0 \\
            0 &0&0&\frac{4!}{dx^4}
        \end{pmatrix}
\end{equation}

\item  $\phi_i=\left[x_{i-2},x_{i-1},x_{i},x_{i+1},x_{i+2}\right]$
This corresponds to the node \( x_i \), associated with all central nodes.

\begin{equation}
\begin{pmatrix}
    1&1&1&1&1\\
    -2&-1&0&1&2\\
    4&1&0&1&4\\
    -8&-1&0&1&8\\
    16&1&0&1&16
\end{pmatrix}
\boldsymbol{\alpha}_{i}
=
\begin{pmatrix}
    0&0&0&0\\
    \frac{1}{dx}& 0&0&0\\
    0& \frac{2!}{dx^2}&0&0 \\
    0 &0& \frac{3!}{dx^3}&0 \\
    0 &0&0&\frac{4!}{dx^4}
\end{pmatrix}
\end{equation}

\item  $\phi_{N-1}=\left[x_{N-4},x_{N-3},x_{N-2},x_{N-1},x_{N}\right]$
This corresponds to the node $x_{N-1}$.
\begin{equation}
        \begin{pmatrix}
            1&1&1&1&1\\
            -3&-2&-1&0&1\\
            9&4&1&0&1\\
            -27&-8&-1&0&1\\
            81&16&1&0&1
        \end{pmatrix}
        \boldsymbol{\alpha}_{N-1}
        =
        \begin{pmatrix}
            0&0&0&0\\
            \frac{1}{dx}& 0&0&0\\
            0& \frac{2!}{dx^2}&0&0 \\
            0 &0& \frac{3!}{dx^3}&0 \\
            0 &0&0&\frac{4!}{dx^4}
        \end{pmatrix}
\end{equation}

\item  $\phi_{N}=\left[x_{N-4},x_{N-3},x_{N-2},x_{N-1},x_{N}\right]$
This corresponds to the node \( x_N \) located at the right boundary.

\begin{equation}
        \begin{pmatrix}
            1&1&1&1&1 \\
            -4&-3&-2&-1&0 \\
            16&9&4&1&0\\
            -64&-27&-8&-1&0\\
            256&81&16&1&0
        \end{pmatrix}
        \boldsymbol{\alpha}_{N}
        =
        \begin{pmatrix}
            0&0&0&0\\
            \frac{1}{dx}& 0&0&0\\
            0& \frac{2!}{dx^2}&0&0 \\
            0 &0& \frac{3!}{dx^3}&0 \\
            0 &0&0&\frac{4!}{dx^4}
        \end{pmatrix}
        \label{eq:ecuaciones_5p}
\end{equation}

\end{enumerate}
Each column of \( \mathbf{f}^{\left(n\right)} \) corresponds to a different derivative order, from \( n=1 \) to \( n=4 \).
The error shown in the equation \eqref{eq:error5}, corresponds to the stencil $\phi_i$. For the remaining stencils we have the following remainders.

\begin{flalign}
    \left| R_{0}\left(\boldsymbol{\alpha}_{0}, \Delta x, f^{\left(5\right)}\right) \right| 
        &\leq \frac{\left| f^{\left(5\right)}\left(c\right) \right| \left( 4\Delta x \right)^{5}}{5!} 
        \left( \left| \alpha_{1}^{0} \right| + \left| \alpha_{2}^{0} \right| + \left| \alpha_{3}^{0} \right| + \left| \alpha_{4}^{0} \right| \right),\quad c\in \left[ x_0,x_4 \right]\\
    \left| R_{1}\left(\boldsymbol{\alpha}_{1}, \Delta x, f^{\left(5\right)}\right) \right| 
        &\leq \frac{\left| f^{\left(5\right)}\left(c\right) \right| \left( 3\Delta x \right)^{5}}{5!} 
        \left( \left| \alpha_{0}^{1} \right| + \left| \alpha_{2}^{1} \right| + \left| \alpha_{3}^{1} \right| + \left| \alpha_{4}^{1} \right| \right) ,\quad c\in \left[ x_0,x_4 \right] \\
    \left| R_{N-1}\left(\boldsymbol{\alpha}_{N-1}, \Delta x, f^{\left(5\right)}\right) \right| 
        &\leq \frac{\left| f^{\left(5\right)}\left(c\right) \right| \left( 3\Delta x \right)^{5}}{5!} 
        \left( \left| \alpha_{0}^{N-1} \right| + \left| \alpha_{1}^{N-1} \right| + \left| \alpha_{2}^{N-1} \right| + \left| \alpha_{4}^{N-1} \right| \right),\quad c\in \left[ x_{N-4},x_N \right]\\
    \left| R_{N}\left(\boldsymbol{\alpha}_{N}, \Delta x, f^{\left(5\right)}\right) \right| 
        &\leq \frac{\left| f^{\left(5\right)}\left(c\right) \right| \left( 4\Delta x \right)^{5}}{5!} 
        \left( \left| \alpha_{0}^{N} \right| + \left| \alpha_{1}^{N} \right| + \left| \alpha_{2}^{N} \right| + \left| \alpha_{3}^{N} \right| \right),\quad c\in \left[ x_{N-4},x_N \right]
\end{flalign}
and the relation with the spatial resolution given the desire error in the approximation 
\begin{align}
        \Delta x &\leq \frac{1}{4} \sqrt[5]{\frac{120\varepsilon}{\left| f^{\left(5\right)}\left(c\right) \right| 
        \left( \left| \alpha_{1}^{0} \right| + \left| \alpha_{2}^{0} \right| + \left| \alpha_{3}^{0} \right| + \left| \alpha_{4}^{0} \right| \right)}}\\
        \Delta x &\leq \frac{1}{3} \sqrt[5]{\frac{120\varepsilon}{\left| f^{\left(5\right)}\left(c\right) \right| 
        \left( \left| \alpha_{0}^{1} \right| + \left| \alpha_{2}^{1} \right| + \left| \alpha_{3}^{1} \right| + \left| \alpha_{4}^{1} \right| \right)}}\\
        \Delta x &\leq \frac{1}{3} \sqrt[5]{\frac{120\varepsilon}{\left| f^{\left(5\right)}\left(c\right) \right| 
        \left( \left| \alpha_{0}^{N-1} \right| + \left| \alpha_{1}^{N-1} \right| + \left| \alpha_{2}^{N-1} \right| + \left| \alpha_{4}^{N-1} \right| \right)}}\\
        \Delta x &\leq \frac{1}{4} \sqrt[5]{\frac{120\varepsilon}{\left| f^{\left(5\right)}\left(c\right) \right| 
        \left( \left| \alpha_{0}^{N} \right| + \left| \alpha_{1}^{N} \right| + \left| \alpha_{2}^{N} \right| + \left| \alpha_{3}^{N} \right| \right)}}
\end{align}
The  challenge in determining the accuracy of the approximation based on the spatial numerical resolution lies in the 5th-order derivative of the function. This challenge arises because, in most cases, the form of the 5th derivative, or even the function itself, is unknown.

\subsection{Forward Model}
In this section, we describe the forward model for each scenario.

\subsubsection{Forward Model for the High-Contrast Stratified Medium}
In \ref{sub:high_contrast_medium}, we outline the model for the wave equation in a homogeneous layer using the wave superposition principle. For the high-contrast stratified medium, we are considering that our medium is the layers composition, each of them homogeneous, and define the solution as the sum of the solution within each of the region.

 Let be $S$ a continuous 1D space domain, and let be $\left\{s_i\right\}$ a partition of $S$ where the intersection of each continuous partition is only in the common boundary. So $S=\cup_{i \in \mathcal{I}}s_i$ and $s_i \cap s_{i+1}=\left\{\partial s_i^{+}\equiv \partial s_{i+1}^{-}\right\}$. So we can express the solution in the stratified medium as the sum of the solution for each of the partitions.

To preserve the elastic properties of the medium, certain conditions have to be taken into consideration. These conditions indicate how the wave acts in the common boundary between the partitions. In \citep[Section 2.2.3]{stein2009introduction}, the authors explained the interaction in the border of each partition in one way in terms of the density properties of the medium $\rho$ and the velocity $c$. These expressions can be translated to only use the velocity term as follows. Let the partitions $s_i$ and $s_{i+1}$, and let $A$ the amplitude of the incident wave from $s_i$ to $s_{i+1}$, $B$ the amplitude of the reflected wave and $C$ the amplitude of the transmitted wave. Then the reflection $R_{i,i+1}$, $R_{i+1,i}$ and transmission $T_{i,i+1}$, $T_{i+1,i}$  coefficients are given by
\begin{equation}
    \begin{aligned}
        R_{i,i+1} & =\frac{c_{i+1}-c_{i}}{c_{i}+c_{i+1}} & T_{i,i+1} & =\frac{2c_{i+1}}{c_{i}+c_{i+1}} \\
        R_{i+1,i} & =-R_{i,i+1}                          & T_{i+1,i} & =2-T_{i,i+1}
        \label{eq:refelxion_transmision}
    \end{aligned}
\end{equation}
here \(R_{i,j}\) and \(T_{i,j}\) stand for the reflextion and transmition of waves traveling from $s_i$ to $s_j$. The equation \eqref{eq:refelxion_transmision} allows calculate the coefficients required to describe the behavior of the wave in the boundary. Each of the equation system \eqref{eq:DF1DWE_ur} and \eqref{eq:DF1DWE_ul} associated to $u_{{s_{i}},r}\left(x,t\right)$ and $u_{{s_{i}},l}\left(x,t\right)$ respectively, has the relation in the boundary condition given by.
\begin{equation}
    \begin{aligned}
        F_{s_i,r}\left(t\right)     & =R_{i,i+1}u_{{s_{i}},r}\left(L_{s_i},t\right)+T_{i+1,i}u_{{s_{i+1}},l}\left(0_{s_{i+1}},t\right) \\
        F_{s_{i+1},l}\left(t\right) & =R_{i+1,i}u_{{s_{i+1}},l}\left(0_{s_i},t\right)+T_{i,i+1}u_{{s_{i}},r}\left(L_{s_{i}},t\right)
    \end{aligned}
    \label{eq:source_at_boundaries}
\end{equation}

To solve the wave equation within the homogeneous layers, we transform the second-order partial differential systems \eqref{eq:DF1DWE_ur} and \eqref{eq:DF1DWE_ul} into a first-order  system.

Let 
\begin{equation}
\boldsymbol{w} = \begin{bmatrix} u \\ \partial_t u \end{bmatrix}.
\end{equation}
Then the system is given by:

\begin{equation}
    \partial_t \boldsymbol{w} = \begin{pmatrix}
        0 & 1 \\
        c^2 \boldsymbol{L}_5\left( \cdot \right) & 0
    \end{pmatrix} \boldsymbol{w} + \boldsymbol{F},
\end{equation}
where \(\boldsymbol{L}_5\) is the five-point second-order operator, and \(\boldsymbol{F}\) is the source term associated with each system.

Using the superposition principle and the equations in \eqref{eq:source_at_boundaries}, we derive the solution to the wave equation in the high-contrast stratified medium.

\subsubsection{Forward Model for the Low-Contrast Stratified Medium}

Hyperbolic systems are highly sensitive to variations in spatial parameters. For the wave equation \eqref{eq:DF1DWE_non-homogeneous}, the parameter in question is the velocity field within the medium. When the velocity field is described using a stepwise function with sharp jumps, it is recommended to avoid using the divergence theorem in numerical approximations, as this approach introduces numerical errors and lacks a clear physical interpretation of the flux \citep{langtangen2016finite,langtangen2016primer,langtangen2017finite}.

In our low-contrast scenario, however, the velocity field is described using a smooth function. This ensures that the divergence theorem remains reliable, as smooth transitions minimize numerical errors and preserve the physical interpretation of the flux. Additionally, we employ the five-point operator to approximate derivatives and use the equations in \eqref{eq:refelxion_transmision} to replicate the behavior of the wave at the boundaries. This combination enhances the precision of the numerical implementation. Hence, our numerical approach is given by

\begin{equation}
    \partial_t \boldsymbol{w} = \begin{pmatrix}
        0 & 1 \\
        \boldsymbol{D}_5 c^2\left( x \right) \cdot \boldsymbol{D}_5 \left( \cdot \right) +c^2\left( x \right)\boldsymbol{L}_5 \left( \cdot \right) & 0
    \end{pmatrix} \boldsymbol{w} + \boldsymbol{F},
\end{equation}
where \(\boldsymbol{D}_5\) is the five-point first-order operator

\subsection{Stein Variational Gradient Descent}
The algorithm proposed in \citep{liu2016stein} provides a strategy to transform a random sample from an initial distribution into a random sample of the target distribution. To do this, they use Stein's operator to approximate the KL divergence between the initial and target distributions. However, applying this algorithm to a parameter identification associated with a differential equation requires more attention to the parameter space support. In this section, we'll explain how to take advantage of the Stein Variational Gradient Descent (SVGD) proposed by \citep{liu2016stein} in parameter identification problems. First, we'll explain a way to choose the \textit{learning rate} or \textit{stepsize}. Then we take into consideration the support constraints and a way to handle them.

\subsubsection{Learning rate or stepsize}
Selecting the learning rate or step size in a gradient descent method is crucial, as it determines both the speed of convergence and the accuracy of convergence to the solution. In \citep{liu2016stein}, the author explains the role of the step size in approximating the target distribution based on a linear transformation from one test distribution to another using the map:
\begin{equation}
T(x) = x + \alpha \phi(x),
\end{equation}
where \(\phi(x)\) represents the direction of the perturbation, and \(\alpha\) denotes the magnitude of the perturbation. When the \(\left|\alpha\right|\) is sufficientelly small, the Jacobian of \(T\) is a full rank and hence
\(T\) is guaranteed to be an one-to-one map by the inverse function theorem. The direction of the perturbation is computed as indicated in \citep{liu2016stein,liu2016kernelized}

\begin{equation}
    \phi(x)=\frac{1}{n} \sum_{j=1}^n\left[k\left(x_j^{\ell}, x\right) \nabla_{x_j^{\ell}} \log p\left(x_j^{\ell}\right)+\nabla_{x_j^{\ell}} k\left(x_j^{\ell}, x\right)\right]
    \label{eq:stein_gradient}
\end{equation}
$k(\cdot,\cdot)$ stands for the Reproducing Kernel Hilbert Spaces \citep{berlinet2011reproducing}

To select the step size, we can choose from several methods. In \citep{ruder2016overview}, the author provides an extensive description of these methods for this purpose. The author discusses their benefits and limitations and compares their performance across various machine learning scenarios.

One of the methods outlined in \citep{ruder2016overview} is the ADAM method \citep{kingma2014adam}, which combines the advantages of two popular optimization techniques: AdaGrad and RMSProp. ADAM computes adaptive learning rates for each parameter by maintaining running averages of both the gradient and its squared value. This approach ensures fast convergence while being robust to noisy gradients and sparse data. Using this method for the learning rate, the SVGD algorithm proposed by \citep{liu2016stein} is given by Algorithm \ref{alg:liu2016stein}.

\begin{algorithm}
    \caption{Bayesian Inference via Variational Gradient Descent}
    \begin{algorithmic}[1]
    \State \textbf{Input:} A target distribution with density function \(p(x)\) and a set of initial particles \(\{x_i^0\}_{i=1}^n\).
    \State \textbf{Output:} A set of particles \(\{x_i\}_{i=1}^n\) that approximates the target distribution \(p(x)\).
    
    \For{\textbf{iteration} \(\ell\)}
        \State \(x_i^{\ell+1} \gets x_i^\ell + \varepsilon_{\ell} \hat{\Phi}^*(x_i^\ell)\), where
        \[
        \hat{\Phi}^*(x) = \frac{1}{n} \sum_{j=1}^n \left[ k(x_j^\ell, x) \nabla_{x_j} \log p(x_j^\ell) + \nabla_{x_j} k(x_j^\ell, x) \right],
        \]
        \Statex \hspace{3.5em} and \(\varepsilon_{\ell}\) is the step size at the \(\ell\)-th iteration.
    \EndFor
    \end{algorithmic}
    \label{alg:liu2016stein}
\end{algorithm}
To determine the learning rate, our approach was to find a suboptimal solution  through an optimization process. We solved the problem defined by :

\begin{equation}
    \alpha^* = \arg \min_{0 \leq \alpha \leq \xi} \mathcal{L}\left( \theta + \alpha \phi\left(\theta\right) \right),
    \label{eq:}
\end{equation}
 where \(\mathcal{L}\) is a loss function to be determined, \(\phi\) represents the direction of perturbation \eqref{eq:stein_gradient}, \(k\left( x,x' \right)=\exp\left(-\frac{1}{2h^2}\left|\left|x-x'\right|\right|_{2}^2\right)\) is a Radial Basis Function(RBF) kernel in the Stein class for smooth density supported in \(\mathbb{R}^n\) \citep{liu2016kernelized}, and \(\xi\) is the maximum step size in that direction. This constraint is what makes the optimization suboptimal.  With this approach, we aim to accelerate convergence while incorporating an adaptive step size.

\subsection{The Evidence Lower Bound}
The Evidence Lower Bound (ELBO) quantifies the proximity of a test distribution \(q\) to a target distribution \(p\). It is defined as:
\begin{equation}
    \text{ELBO}(q) = \mathbb{E}[\log p(z)] + \mathbb{E}[\log p(x \mid z)] - \mathbb{E}[\log q(z)]
\end{equation}
which can be rewritten as:
\begin{align}
    \text{ELBO}(q) &= \mathbb{E}[\log p(x \mid z)] - \text{KL}(q(z) \| p(z)), 
    \label{eq:elbo_definition}
\end{align}
where the expectations are taken over the test distribution \(q\). Here, \(z\) represents the parameter vector, \(x\) denotes the observation vector, and \(\text{KL}(\cdot)\) refers to the Kullback-Leibler (KL) divergence, defined as:
\begin{equation}
\text{KL}(q \| p) = \frac{1}{2} \left[ \log\frac{\det\Sigma_p}{\det\Sigma_q} - d + \text{Tr}(\Sigma_p^{-1} \Sigma_q) + (\mu_p - \mu_q)^\top \Sigma_p^{-1} (\mu_p - \mu_q) \right],
\end{equation}
where \(\mu_p, \mu_q\) are the means, and \(\Sigma_p, \Sigma_q\) are the covariance matrices of \(p\) and \(q\), respectively.

The term \(\mathbb{E}[\log p(x \mid z)]\) can be approximated with MCMC integration as:

\begin{equation}
    \mathbb{E}[\log p(x \mid z)] = \int \log p(x \mid z) \, q(z) \, dz \approx \frac{1}{N} \sum_{i=1}^N \log p(x \mid z_i), \quad z_i \sim q(z)
\end{equation}
where \(N\) represents the number of samples drawn from \(q(z)\). Using this approximation, the ELBO can be expressed as:
\begin{equation}
    \text{ELBO}(q) \approx \frac{1}{N} \sum_{i=1}^N \log p(x \mid z_i) - \frac{1}{2} \left[ \log\frac{\det\Sigma_p}{\det\Sigma_q} - d + \text{Tr}(\Sigma_p^{-1} \Sigma_q) + (\mu_p - \mu_q)^\top \Sigma_p^{-1} (\mu_p - \mu_q) \right].
\end{equation}

The ELBO is widely employed in variational inference as an alternative to directly minimizing the KL divergence. From Equation \eqref{eq:elbo_definition}, maximizing the ELBO is equivalent to minimizing the KL divergence.

The use of the Evidence Lower Bound (ELBO) for approximating a target distribution was extensively explored in \citep{blei2017variational}. In this work, the author employed a parametric family of distributions to approximate the target density. This method assumed that each parameter was independently distributed, neglecting the underlying relationships between parameters. Consequently, this approach increased the dimensionality of the parameter space by incorporating hyperparameters of the chosen parametric family.

In contrast, our approach focuses on providing a Gaussian approximation of the target distribution centered around its local modes. To achieve this, we introduced a novel loss function tailored to capture the proximity of the approximation to the target distribution while preserving computational efficiency.

\subsection{Loss Function}



The goal of optimization is to reduce the total loss across the complete dataset, ensuring that the model predictions align more closely with the true values.


In the gradient descent method, the loss function is minimized iteratively by adjusting the model parameters. The gradient of the loss function with respect to the parameters determines the direction of the steepest ascent. By updating parameters in the opposite direction, gradient descent progressively reduces the loss function value. 



In variational inference, the loss function is represented by the negative ELBO. The ELBO measures the alignment between the approximate posterior distribution \( q \) and the true posterior of model parameters. By maximizing the ELBO, the framework ensures a close approximation to the true posterior by balancing: \textit{the likelihood of observed data} and \textit{a regularization term that imposes prior constraints}



Building upon these foundations, the proposed loss function integrates the benefits of gradient descent, the weighted gradient introduced by \citep{liu2016stein}, and the ELBO. 

\begin{equation}
    \mathcal{L}\left( \Theta, q \right) = (1-\omega)\left\|\nabla_{x}\log\left( \Theta, \boldsymbol{y} \right)\right\|_{\infty} - \omega \, \text{ELBO}\left( q \right),
    \label{eq:loss_function}
\end{equation}
where: \( \Theta \) is a set of particles, \( \boldsymbol{y} \) are the observation, \( q \) is the test distribution,  \( \omega \) a weight parameter, and  \( \|\cdot\|_{\infty} \) is the supremum norm defined as:
\begin{equation}
\left\|\nabla_{x}\log\left( \Theta, \boldsymbol{y} \right)\right\|_{\infty} = \sup\left\{\left|\nabla_{x}\log\left( \theta_i, \boldsymbol{y} \right)\right| : \theta_i \in \Theta \right\}.
\end{equation}


The parameter \( \omega \) governs the trade-off between gradient-driven optimization and posterior alignment:
\begin{itemize}
    \item \textbf{\( \omega = 0 \)}: Focuses solely on minimizing the supremum norm, driving particles toward local optima.
    \item \textbf{\( \omega = 1 \)}: Emphasizes maximizing the ELBO, aligning the test distribution \( q \) with the target.
    \item \textbf{\( \omega \in \left( 0,1 \right) \)}: Balances the two objectives, ensuring both stability and accuracy.
    \item \textbf{\( \omega \notin \left[ 0,1 \right] \)}: The loss function becomes ill-defined, as it simultaneously minimizes and maximizes conflicting objectives.
\end{itemize}


Minimizing the loss function with respect to \( \omega \) yields:
\begin{equation}
\nabla_{\omega}\mathcal{L}\left( \Theta, q \right) = -\left\|\nabla_{x}\log\left( \Theta, \boldsymbol{y} \right)\right\|_{\infty} - \text{ELBO}\left( q \right).
\end{equation}
the condition for achieving the minimum is:
\begin{equation}
    \left\|\nabla_{x}\log\left( \Theta, \boldsymbol{y} \right)\right\|_{\infty} = - \text{ELBO}\left( q \right).
    \label{eq:loss_function_minimum_necessary_condition}
\end{equation}
This result ensures robust optimization by balancing the contributions of gradient evaluation and posterior approximation. The proposed loss function achieves convergence irrespective of the specific choice of \( \omega \), provided the stated condition holds. Moreover, the minimum of the loss function lies within the hyperplane formed by the parameter space \(\Omega\) and the space of the test distribution \(\mathcal{Q}\), and the mínimum value for \(\mathcal{L}\) is achieve when the relation in equation \eqref{eq:loss_function_minimum_necessary_condition} holds.

\subsection{Liu's Method and Learning Rate}

In \citep{liu2016stein} the author selects the learning rate to be sufficiently small to ensure invertible transformations between consecutive states while minimizing the KL divergence between the test function \( q \) and the target distribution \( p \).

Given a sample at state \(\ell+1\), \(\Theta^{\ell+1}\), Algorithm \ref{alg:liu2016stein} guarantees that \( q_{\ell+1}(x) = q_{\ell \left[T_{\ell}\right]}(x) \) reduces the KL divergence by an amount \(\varepsilon_{\ell+1}D^2(q_{\ell+1}, p) = \varepsilon_{\ell+1}\left\|\phi_{(q_{\ell+1}, p)}(x)\right\|\) \citep{liu2016stein}. According to equation \eqref{eq:elbo_definition}, this reduction is equivalent to increasing the -ELBO of \( q_{\ell+1} \) by the same magnitude. Substituting into \eqref{eq:loss_function}, we obtain:
\begin{equation}
    (1-\omega)\left\|\nabla_{x}\log\left( \Theta^{\ell+1}, \boldsymbol{y} \right)\right\|_{\infty}-w\text{ELBO}(q_{\ell+1}) - w\varepsilon_{\ell+1}D^2(q_{\ell+1}, p) \leq (1-\omega)\left\|\nabla_{x}\log\left( \Theta^{\ell+1}, \boldsymbol{y} \right)\right\|_{\infty}-w(\text{ELBO}(q_{\ell+1})).
\end{equation}

This result demonstrates that the sequence of values \(\{\varepsilon_{\ell}\}\) proposed in \citep{liu2016stein} minimizes the proposed loss function.

In our approach, we consider a suboptimal learning rate (not necessarily small) along the gradient direction that minimizes the loss function \(\mathcal{L}\). As shown in \eqref{eq:loss_function_minimum_necessary_condition}, the minimum of the loss function does not depend on the value of the free parameter \( w \). Instead, the minimum depends on the descent trajectory, which is determined solely by the sequence of values \(\{\alpha_{\ell}\}\) and the weighted gradient $\phi$. The suboptimal value in our optimization process depends on the admissible interval for each \(\alpha_{\ell}\), denoted as \(\xi_{\ell}\). This approach makes the descent with invertible transformations a specific case of the minimization process for the loss function \(\mathcal{L}\).

\begin{figure}
    \centering
    \includegraphics{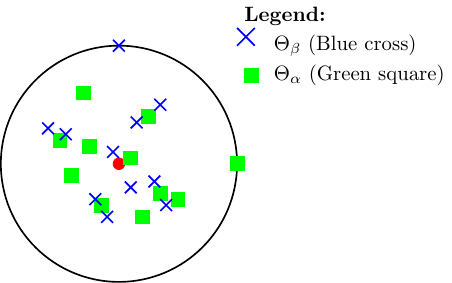}
    \caption{The figure shows the final state of the approximation for two sequences of learning rates. Particles are distributed in space based on their parameter values, and the distance to the center corresponds to the magnitude of their gradient. The largest gradient magnitude, \(\varsigma\), is associated with particles at the boundary.}
    \label{fig:circle_with_random_samples_legend}
\end{figure}

Let \(\Theta^0\) denote the initial set of particles, \(q\) the test distribution, \(\left\{\alpha_i\right\}_{i=1}^{n_{\alpha}}\) and \(\left\{\beta_i\right\}_{i=1}^{n_{\beta}}\) two distinct sequences of learning rates convergin to the minimum of $\mathcal{L}$, and \(\varsigma = \min \mathcal{L}\). From \eqref{eq:loss_function_minimum_necessary_condition}, we have 
\begin{equation}
    \left\|\nabla_{x}\log\left( \Theta, \boldsymbol{y} \right)\right\|_{\infty} = \varsigma,
\end{equation}
which implies that the final states of the samples, \(\Theta_{\alpha}\) and \(\Theta_{\beta}\), are not necessarily the same. See Figure \ref{fig:circle_with_random_samples_legend} for an example of two final states for the same initial set of particles.

\subsection{Greedy Stein Variational Gradient Descent}
The transitions between states in the particle set are substantially influenced by both the step size and the direction indicated by the weighted gradient, denoted as $\phi$. The justification for selecting the value of $\alpha$ is to ensure that its absolute values remain sufficiently small, thus preserving the invertibility of the linear transformation from one state to another.

This proposal capitalizes on the weighted gradient while simultaneously increasing the step size. The magnitude of the step size is utilized in the optimization problem, recognizing that a suboptimal approximation of this parameter is employed to minimize the loss function \eqref{eq:loss_function}. This highlights the importance of the loss function as a central element of this investigation.

The testing function may be selected from a diverse array of parametric and non-parametric distributions. The ELBO serves to ensure the closeness of this function to the posterior distribution. Our methodology involves providing a Gaussian approximation centered on the local modes of each particle. Although the supremum norm aids in this aspect, it is insufficient in determining the form of the distribution. To effectively address this limitation, we employ Kernel Density Estimation (KDE), which offers the desired type of approximation. The following Algortihm \ref{alg:GSVGD} incorportates all these aspects.

\begin{algorithm}
    \caption{Greedy Stein Variational Gradient Descent for approximating  the posterior distribution}
    \begin{algorithmic}[1]
    \State \textbf{Input:} Initial set of particles \(\Theta^0 = \{\boldsymbol{\theta}^0_i\}_{i=1}^m\).
    \State \textbf{Output:} A set of particles \(\Theta^\ell = \{\boldsymbol{\theta}^\ell_i\}_{i=1}^m\) that minimizes the loss function \(\mathcal{L}\).

    \For{\textbf{iteration} \(\ell = 0\) to \(n\)}
        \State Compute the test distribution \(q^\ell(\boldsymbol{\theta}) \gets \operatorname{KDE}_{\Theta^\ell, \boldsymbol{\omega}}(\boldsymbol{\theta})\).
        \If{\(\mathcal{L}(\Theta^\ell, q^\ell) > \epsilon\)}
            \State Define the candidate set of particles:
            \[
            \Phi(\alpha) \gets \{\boldsymbol{\theta}^\ell_i + \alpha \phi(\boldsymbol{\theta}^\ell_i)\}_{i=1}^m,
            \]
            \Statex \hspace{1.5em} where \(\phi\) is the gradient-based update direction.
            \State Compute the candidate test distribution:
            \[
            q^\alpha(\boldsymbol{\theta}) \gets \operatorname{KDE}_{\Phi, \boldsymbol{\omega}}(\boldsymbol{\theta}).
            \]
            \State Find the optimal step size:
            \[
            \alpha_\ell \gets \arg \min_{\alpha \in (0,\xi)} \mathcal{L}(\Phi, q^\alpha).
            \]
            \State Update the set of particles:
            \[
            \Theta^{\ell+1} \gets \Phi(\alpha_\ell).
            \]
        \Else
            \State \Return \(\Theta^\ell\)
        \EndIf
    \EndFor
    \State \Return \(\Theta^\ell\)
    \end{algorithmic}
    \label{alg:GSVGD}
\end{algorithm}

\subsubsection{Explanation of the Algorithm}
This algorithm effectively integrates the weighted gradient, step size optimization, and posterior approximation, ensuring convergence to a Gaussain approximation of the posterior distribution around its local modes within a computationally efficient framework.
\begin{enumerate}
    \item \textbf{Initialization:} The algorithm begins with an initial particle set \(\Theta^0\).
    \item \textbf{Test Distribution:} At each iteration, the test distribution is computed using KDE.
    \item \textbf{Loss Evaluation:} If the current loss \(\mathcal{L}(\Theta^\ell, q^\ell)\) exceeds the threshold \(\epsilon\), the algorithm proceeds to optimize.
    \item \textbf{Candidate Particles:} A candidate set of particles is defined using the gradient-based update direction \(\phi\).
    \item \textbf{Step Size Optimization:} The step size \(\alpha_\ell\) is optimized to minimize the loss for the candidate test distribution.
    \item \textbf{Particle Update:} The particle set is updated using the optimized step size.
    \item \textbf{Convergence:} If the loss is below the threshold, the algorithm terminates and returns the current particle set.
\end{enumerate}

The primary bottleneck of the algorithm lies in the optimization process. Ideally, a full optimization procedure that minimizes the loss function \(\mathcal{L}\) would be performed at each step. However, this approach is computationally intensive, as it requires evaluating both the gradient and the Hessian of \(\mathcal{L}\) at every iteration. To address this challenge, we employ a derivative-free optimization method, which mitigates the computational burden while maintaining the efficacy of the algorithm.

\section{Results}\label{sec:results}

This section presents the numerical results of applying the previous methods to two wave prospecting models with similar characteristics. First, we determined the spatial resolution required for the discrete approximation to replicate the continuous physical model accurately. Then, we show the numerical experiments conducted.





\subsection{Discrete Approximation of Fundamental Components of the Physical Model}
For the discrete approximation of the physical model, we identified the key aspects to consider when solving the forward model. The factors analyzed include the governing physical model, boundary conditions, the type of wave function to approximate, and the numerical resolution of the discrete model.
The governing physical model is the wave equation with absorbing boundary conditions. Solving this model using the finite difference method involves discretizing the first- and second-order derivatives. Similarly, absorbing boundary conditions require the proper discretization of the first-order derivative.

The approximation of an \( n \)-order derivative and the coefficient values for the five-point operator are obtained from the equations \eqref{eq:5po_discrete_system}

Functions of the form \( f(x \pm ct) \) describe waves that propagate in time and space \citep{halliday2013fundamentals}. We focus on cosine waves and waves with exponential decay, as these two types accurately describe transverse waves \citep{halliday2013fundamentals, langtangen2016finite, igel2017computational, shearer2019introduction}.

Cosine waves are defined as
\begin{equation}
    f(x,t) = A\cos(kx \pm \omega t)
    \label{eq:onda_coseno}
\end{equation}
and exponential decay waves as
\begin{equation}
    f(x,t) = \frac{1}{\sqrt{2\pi}\sigma}\exp\left(-\frac{|x \pm ct|^2}{2\sigma^2}\right)
    \label{eq:onda_exponencial}
\end{equation}

\begin{figure}[H]
    \centering
    \begin{minipage}[b]{0.7\textheight}
        
        \includegraphics[width=\linewidth]{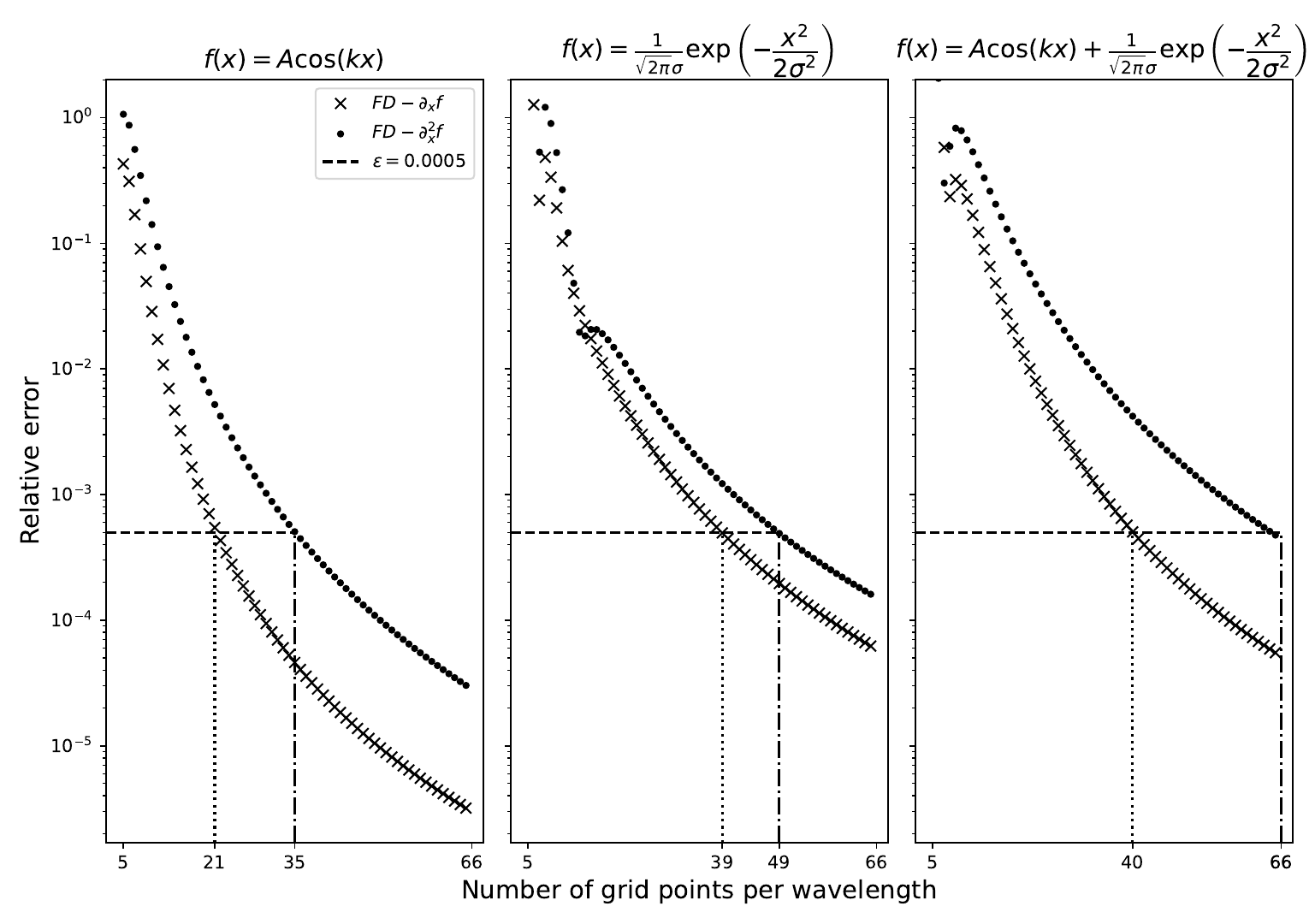}
        \caption{The figure shows the minimum number of grid points per wavelength required to approximate the first and second order derivatives using \textit{five-point operator} with a relative error less than or equal to $0.0005\left(0.05\%\right)$.}
        \label{fig:relative_error} 
    \end{minipage}
    
\end{figure}

\begin{figure}[H]
    \centering
    \begin{minipage}[b]{0.7\textheight}
    
    \includegraphics[width=\linewidth]{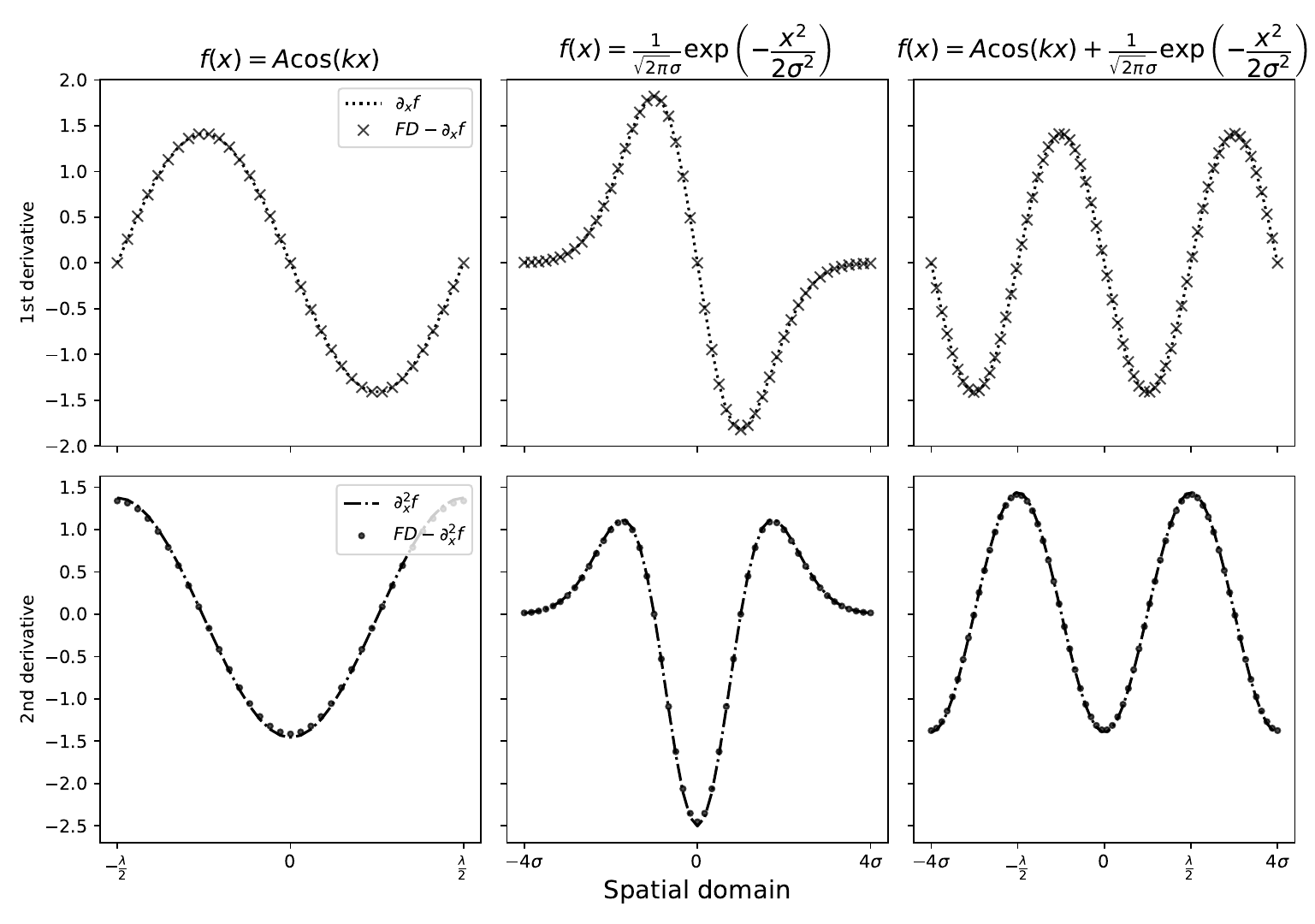}
    \caption{The figure shows the first and second-order derivative reconstruction from the \textit{five-point operator}. From left to right used 35, 49, and 55 grid points per wavelength}
    \label{fig:derivative_approximation}
    \end{minipage}
\end{figure}

We aim to approximate the spatial derivative at each time step and analyze the numerical error using the five-point operator. The waves in \eqref{eq:onda_coseno} and \eqref{eq:onda_exponencial} are traveling waves that retain their shape in a homogeneous medium. We approximate the first- and second-order derivatives of these functions at \( t=0 \), including the sum of both wave types. The experiment increases the number of spatial nodes per wavelength to improve the derivative approximation. Results are shown in Figures \ref{fig:relative_error} and \ref{fig:derivative_approximation}.

\subsubsection{Open boundary approximation}
\begin{figure}[H]
    \centering
    \begin{minipage}[b]{0.6\textheight}
        \includegraphics[width=\textwidth]{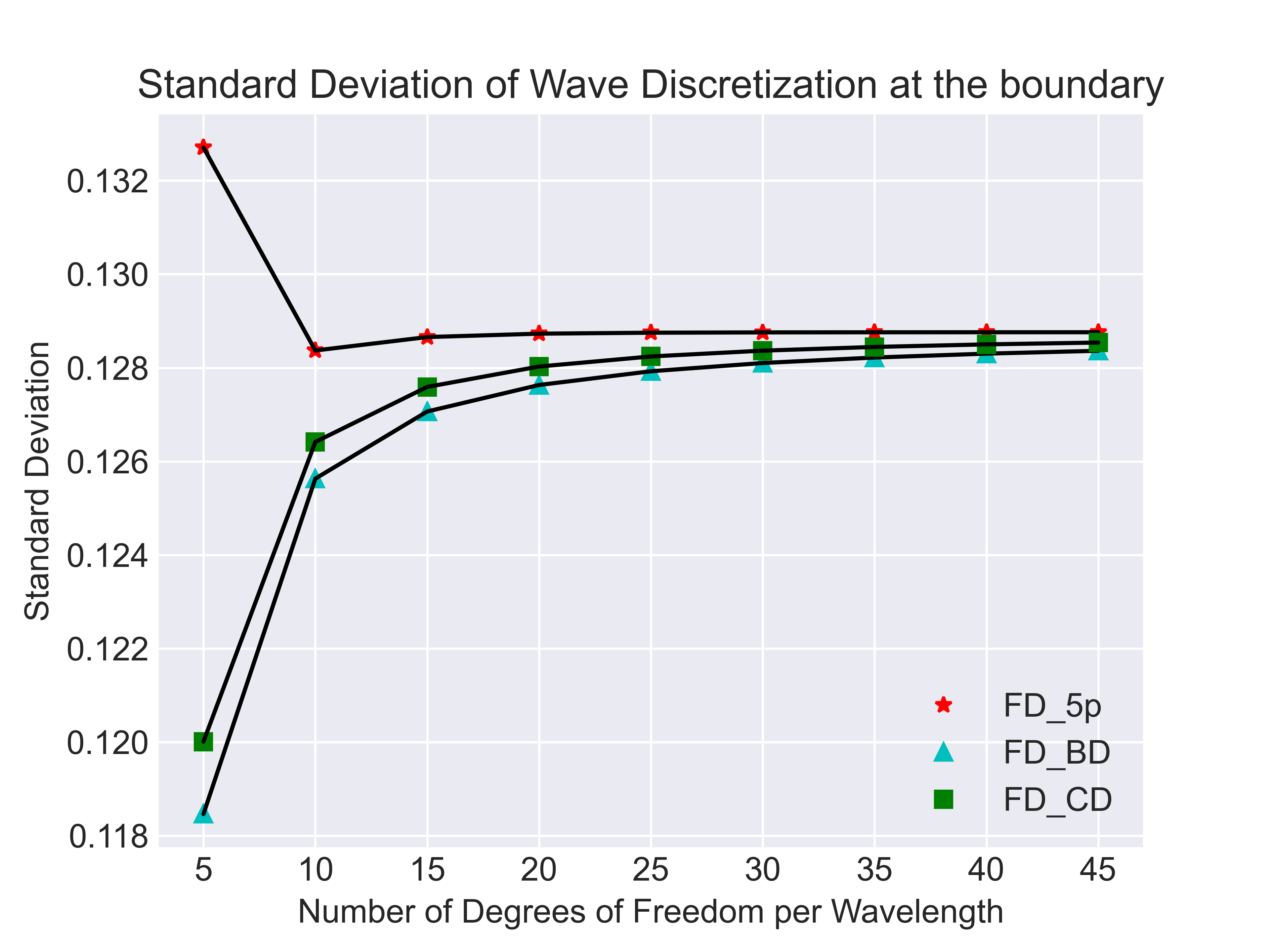}
    \end{minipage}
    \caption{The five-point operator (FD\_5p) requires fewer degrees of freedom to achieve convergence compared to the backward (FD\_BD) and centered (FD\_CD) finite difference operators.}
    \label{fig:comparacion_diferencias_finitas}
\end{figure}

As described previously, our model involves approximating open boundary conditions. Without an analytical solution for direct comparison, we require an alternative approach to evaluate the numerical solutions.
Our objective is to demonstrate the convergence, stability, and consistency of the proposed numerical method. The accuracy of finite difference methods improves as long as the CFL condition is satisfied and the spatial resolution is sufficiently fine. Together, these conditions ensure that each finite difference method converges to the same solution as the number of spatial nodes increases.
This experiment focused on observing wave behavior at the boundary and comparing our results.
Figure \eqref{fig:comparacion_diferencias_finitas} shows the numerical results of the comparison.

\subsection{Prospection models}
\subsubsection{High contrast stratified medium}
\begin{figure}[H]
    \centering
    \includegraphics{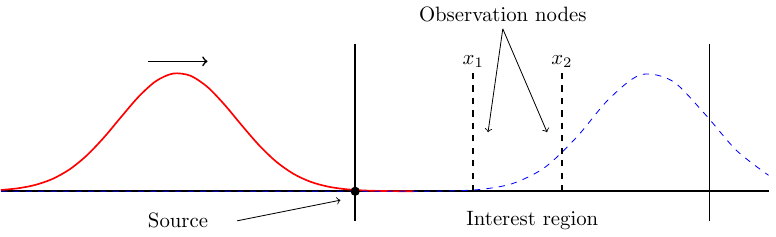}
    \caption{The figure shows a wave entering an area of interest, where it is observed at nodes $x_1$ and $x_2$, and then travels out of the region through the open boundary on the right.}
    \label{fig:diseno_del_experimento}
\end{figure}

We considered a longitudinal bar composed of various materials (to be determined), with several sensors (nodes) placed along it to measure the amplitude of a traveling wave. The wave propagates continuously from the left boundary (source) to the right Figure \ref{fig:diseno_del_experimento}. In the region of interest, we assume the square of the wave speed, \( v^2 \), ranges between 1500 m/s and 3000 m/s.


\begin{figure}[htbp]
    \centering
    \includegraphics[scale=0.9]{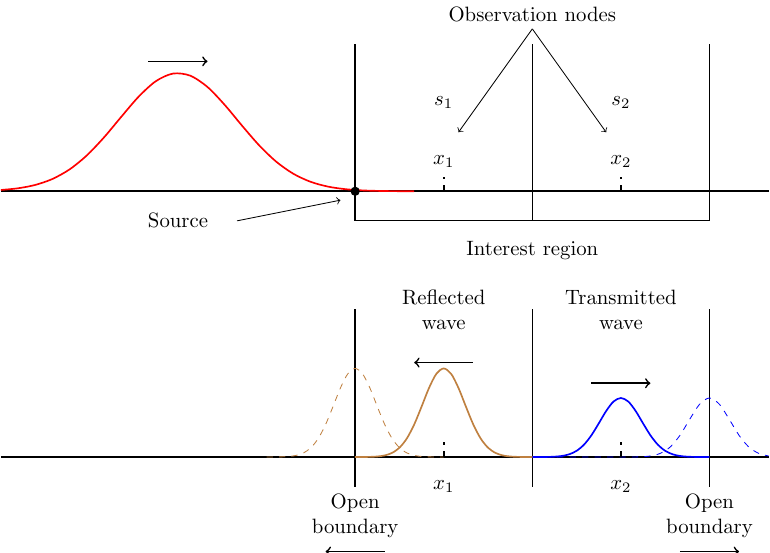} 
    \caption{The top figure illustrates the setup of the surrogate model based on the medium used in the experiment. The bottom figure shows the interaction between regions within the surrogate model.}
    \label{fig:diseno_del_modelo_surrogate}
\end{figure}

In this model, the medium is divided into two regions, each with a constant wave propagation speed. The boundaries between these regions act as sources for the adjacent media. For this reason, we considered the reflection and transmission coefficients previously defined in equation \eqref{eq:refelxion_transmision}, as well as the boundary interactions described in \eqref{eq:source_at_boundaries}. Since each region is homogeneous, we assume a linear wave model where boundaries act as sources for neighboring regions.
Equation \eqref{eq:wave_solution_homogeneous} shows the solution to the wave equation in a homogeneous medium, using superposition principles through systems \eqref{eq:DF1DWE_ur} and \eqref{eq:DF1DWE_ul}. Extending this approach to each region in the surrogate model yields the following general formulation:

\begin{equation}
    \begin{aligned}
        u\left(x,t\right) &= \sum_{i \in \mathcal{I}} u_{s_i}\left(x,t\right) \mathbb{I}_{s_{i}}(x)
        \label{eq:forward_model_equation}
    \end{aligned}
\end{equation}
where
\begin{equation}
    \begin{aligned}
        \mathbb{I}_{s_{i}}(x) & =\begin{cases} 0, & x \not\in \mathcal{X}_{s_{i}} \\ 1, & x \in \mathcal{X}_{s_i} \end{cases}
    \end{aligned}
\end{equation}
Here, \( \mathcal{X}_{s_{i}} \) is the spatial domain for partition \( s_{i} \), and the wave solution in region \( s_{i} \) is given by

\begin{equation}
    u_{s_{i}}\left(x,t\right) = u_{s_{i},r}\left(x,t\right) + u_{s_{i},l}\left(x,t\right)
    \label{eq:wave_solution_homogeneous_s}
\end{equation}

\begin{equation}
    \begin{array}{rlll}
        \partial_{t}^{2}{u_{s_{i},r}} - c_{s_{i}}^2 \partial_{x}^{2}{u_{s_{i},r}} & = 0, & x \in (0, L), & t \in (0, T] \\
        u_{s_{i},r}(x, 0) & = 0, & x \in [0, L] & \\
        \partial_{t}{u_{s_{i},r}}(x, 0) & = 0, & x \in [0, L] & \\
        u_{s_{i},r}(0, t) & = F_{s_{i},l}(t), & t > 0 & \\
        \partial_{t}u_{s_{i},r}(L, t) & = -c_{s_{i}}\partial_{x}u_{s_{i},r}(L, t) & t > 0 &
    \end{array}
    \label{eq:DF1DWE_usr}
\end{equation}

\begin{equation}
    \begin{array}{rlll}
        \partial_{t}{2}{u_{s_{i},l}} - c_{s_{i}}^2 \partial_{x}{2}{u_{s_{i},l}} & = 0, & x \in (0, L), & t \in (0, T] \\
        u_{s_{i},l}(x, 0) & = 0, & x \in [0, L] & \\
        \partial_{t}{u_{s_{i},l}}(x, 0) & = 0, & x \in [0, L] & \\
        \partial_{t}u_{s_{i},l}(0, t) & = c_{s_{i}}\partial_{x}u_{s_{i},l}(0, t) & t > 0 & \\
        u_{s_{i},l}(L, t) & = F_{s_{i},r}(t) & t > 0 &
    \end{array}
    \label{eq:DF1DWE_usl}
\end{equation}
Here, \( u_{s_{i},r} \) and \( u_{s_{i},l} \) represent waves traveling from left-to-right and right-to-left, respectively, and \( s_{i} \) denotes the region in which they propagate. The source term \( f \) has been omitted because each boundary acts as a source term for the system. See Figure \ref{fig:diseno_del_modelo_surrogate}


\subsubsection{Bayesian Model}

Given \( s_{i} \), let \( \mathbb{X}_{s_{i}} \subseteq \mathcal{X}_{s_{i}} \) represent the set of spatial nodes where observations are made, and let \( u_{s_{i}} \) denote the solution of the forward model in that region, as described by equation \eqref{eq:forward_model_equation}. The vector of observations is denoted by \( \mathbf{y}_{s_{i}}\left(t\right) = \left(y_{s_{i},1}\left(t\right), \dots, y_{s_{i},n_{s_{i}}}\left(t\right)\right) \), with \( n_{s_{i}} = \vert \mathbb{X}_{s_{i}} \vert \), representing the number of observations. The additive model is given by:

\begin{equation}
    \begin{aligned}
        \mathbf{y}_{s_{i}}\left(t\right) &= u_{s_{i}}\left(\mathbb{X}_{s_{i}},t,\theta_{s_{i}} \right) + \varepsilon_{s_{i}}, \quad \varepsilon_{s_{i}} \sim \mathcal{N}(0,\sigma_{s_{i}^2}\mathbf{I}) \\
        \mathbf{y}_{s_{i}}(t) & \sim \mathcal{N}(u_{s_{i}}\left(\mathbb{X}_{s_{i}},t,\theta_{s_{i}} \right), \sigma_{s_{i}^2} \mathbf{I})
    \end{aligned}
    \label{eq:modelo_bayesiano_modelo_surrogate}
\end{equation}
where the likelihood distribution is defined as follows:
\begin{equation}
    \begin{aligned}
        L(\mathbf{y} \mid \boldsymbol{\theta}) &= \prod_{i=1}^{N} \pi_{s_{i}}(\mathbf{y}_{s_{i}} \mid \theta_{s_{i}}) \\
        \pi_{s_{i}}(\mathbf{y}_{s_{i}} \mid \theta_{s_{i}}) &= \frac{1}{\sqrt{\left(2 \pi \sigma_{s_{i}}^{2}\right)^{n_{s_{i}}}}} \exp\left(-\frac{1}{2 \sigma_{s_{i}}^{2}} \Vert \mathbf{y}_{s_{i}}(t) - u_{s_{i}}\left(\mathbb{X}_{s_{i}},t,\theta_{s_{i}} \right) \Vert_{L^2\left[0,T\right]^{n_{s_{i}}}}^{2}\right)
    \end{aligned}
\end{equation}
where \( \theta = \left(\theta_{s_{1}}, \dots, \theta_{s_{N}} \right) \) represents the vector of parameters and \( \mathbf{y} = \left(y_{s_{1}}, \dots, y_{s_{N}}\right) \) represents the vector of observations within the region of interest.

Applying Bayes' theorem,
\begin{equation}
    \begin{aligned}
        \pi(\theta \mid \mathbf{y}) &= \frac{L(\mathbf{y} \mid \theta) \pi(\theta)}{Z} \\
        \log \pi(\theta \mid \mathbf{y}) &= \log L(\mathbf{y} \mid \theta) + \log \pi(\theta) - \log Z
    \end{aligned}
\end{equation}
we obtain an approximation to the log posterior distribution:
\begin{equation}
    \log \pi(\theta \mid \mathbf{y}) \approx \sum_{i=1}^{n} -\frac{1}{2 \sigma_{s_{i}}^{2}} \Vert \mathbf{y}_{s_{i}}(t) - u\left(\mathbb{X}_{s_{i}},t \right) \Vert_{L\left[0,T\right]^{n_{s_{i}}}}^{2} + \left( \theta - \theta_{0} \right)^\intercal \mathbf{M}^{-1} \left( \theta - \theta_{0} \right)
\end{equation}
where
\begin{equation}
    \sum_{i=1}^{N} -\frac{1}{2 \sigma_{s_{i}}^{2}} \Vert \mathbf{y}_{s_{i}}(t) - u\left(\mathbb{X}_{s_{i}},t \right) \Vert_{L\left[0,T\right]^{n_{s_i}}}^{2} = \int_{0}^{T} \sum_{i=1}^{N} -\frac{1}{2 \sigma_{s_{i}}^{2}} \left( \mathbf{y}_{s_{i}}(t) - u\left(\mathbb{X}_{s_{i}},t \right) \right)^{2} \, dt
\end{equation}
and
\begin{equation}
    u\left(x,t\right) = \bigoplus_{i=1}^{N} u_{s_i}\left(x,t\right)
\end{equation}

\subsubsection{Numerical Results}
\begin{figure}[H]
    \centering
    \begin{minipage}[b][\textheight]{\textwidth}
    \subfloat[G-SVGD $\omega = 0$]{
        \label{fig:G-SVGD00}
        \includegraphics[width=0.45\textwidth]{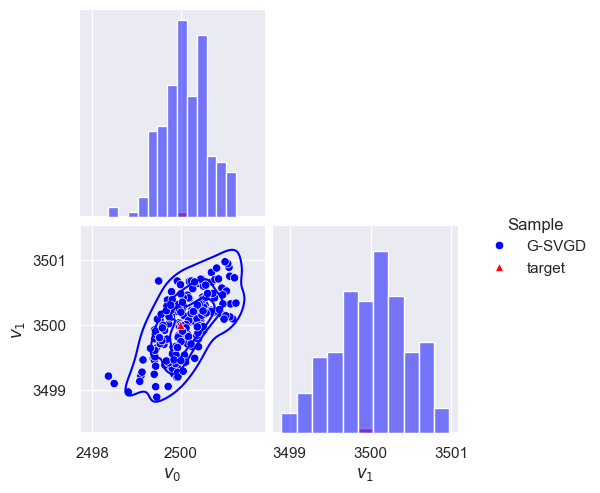}}
    \hfill
    \subfloat[MCMC t-walk]{
        \label{fig:mcmc_hist}
        \includegraphics[width=0.45\textwidth]{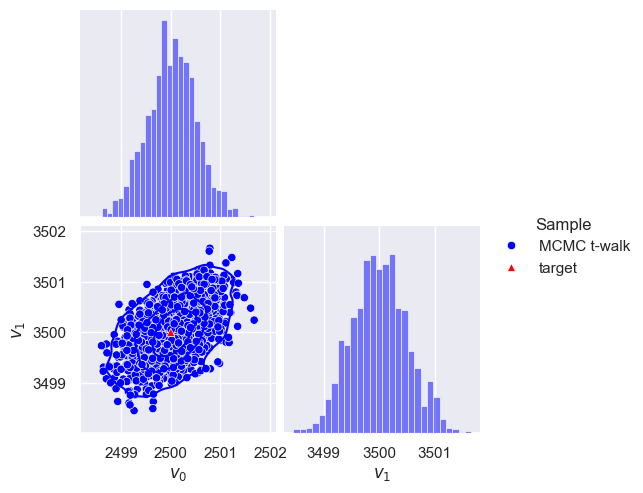}}
    
    \vspace{0.3cm}  
    
    \subfloat[G-SVGD $\omega = 0.5$]{
        \label{fig:G-SVGD05}
        \includegraphics[width=0.45\textwidth]{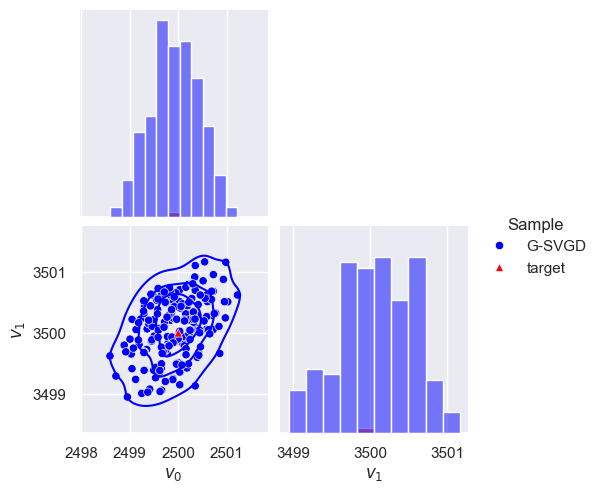}}
    \hfill
    \subfloat[A-SVGD]{
        \label{fig:A-SVGD_hist}
        \includegraphics[width=0.45\textwidth]{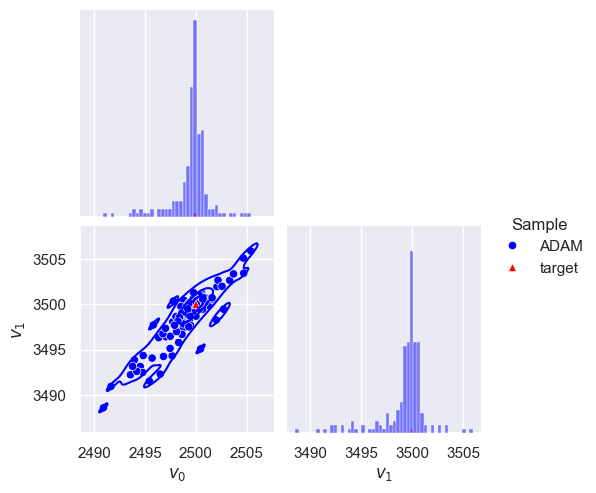}}
    
    \vspace{0.3cm}  

    \subfloat[G-SVGD $\omega = 1$]{
        \label{fig:G-SVGD10}
        \includegraphics[width=0.45\textwidth]{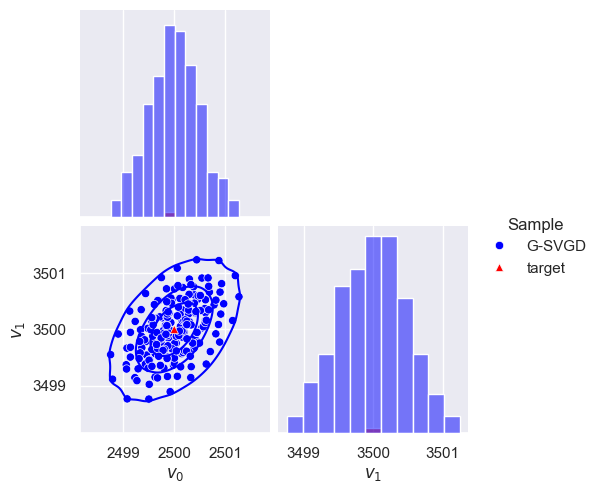}}

    \caption{Histograms of samples obtained using each method for posterior distribution sampling. The red triangle indicates the velocity values used to generate the observations.}
    \label{fig:histograms}
\end{minipage}
\end{figure}

\begin{figure}[H]
    \centering
    \begin{minipage}[b]{0.6\textheight}
        \includegraphics[width=\textwidth]{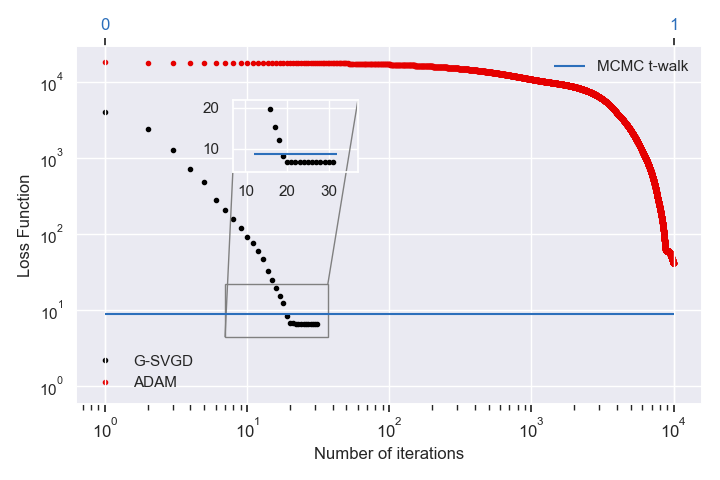}  
        \caption{Convergence comparison between G-SVGD, A-SVGD, and MCMC t-walk for $\omega=0.0$.}
        \label{fig:convergence_w00} 
    \end{minipage}
\end{figure}

\begin{figure}[H]
    \centering
    \begin{minipage}[b]{0.6\textheight}
        \includegraphics[width=\textwidth]{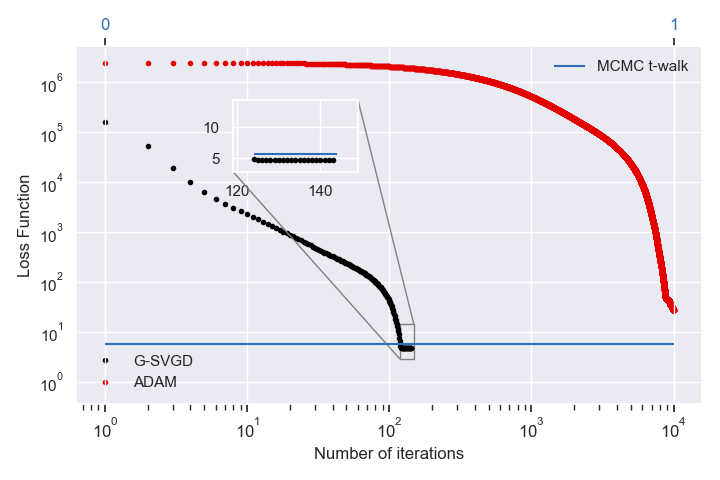}  
        \caption{Convergence comparison between G-SVGD, A-SVGD, and MCMC t-walk for $\omega=0.5$.}
    \label{fig:convergence_w05}
\end{minipage}
\end{figure}

\begin{figure}[H]
    \centering
    \begin{minipage}[b]{0.6\textheight}
        \includegraphics[width=\textwidth]{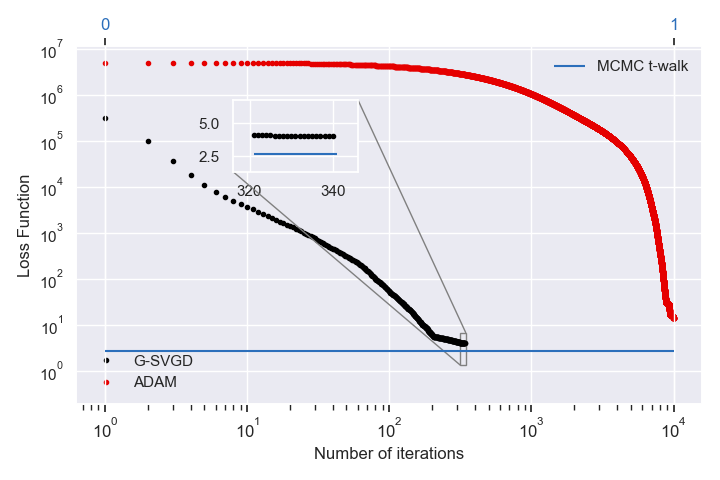}  
        
        \caption{Convergence comparison between G-SVGD, A-SVGD, and MCMC t-walk for $\omega=1.0$.}
    \label{fig:convergence_w10}
\end{minipage}
\end{figure}

Using the proposed loss function in \eqref{eq:loss_function}, we evaluated values of $\omega = \{0, 0.5, 1\}$. These values represent different scenarios achievable with this function. Results were compared with the t-walk method \citep{christen2010general} and the SVGD method, using the ADAM optimizer \citep{kingma2014adam} for learning rate selection in the latter case. Figure \ref{fig:histograms} shows the histograms of the simulations. Figures \ref{fig:convergence_w00}, \ref{fig:convergence_w05}, and \ref{fig:convergence_w10} provide a convergence comparison for each method using the respective loss function.

\subsection{Low contrast stratified medium}

In this experiment, we analyzed a stratified medium with low contrast. The medium design is similar to the one described in Figure \ref{fig:diseno_del_experimento}. It is assumed that the transverse wave propagation speed can be approximated by a smooth curve.

\subsubsection{Forward Mapping}
The wave prospecting model is similar to the system proposed in \eqref{eq:DF1DWE_non-homogeneous}. In this case, interactions within the medium are considered rather than at the boundary, thus maintaining the design illustrated in Figure \eqref{fig:diseno_del_experimento}. The forward model is given by system \eqref{eq:low_contrast}:

\begin{equation}
    \begin{array}{rlll}
        \partial_{t}^{2}{u}\left(x,t\right) -\nabla \left(c^2\left(x\right)\nabla u\left(x,t\right)\right)	& =0, 							& x \in(0, L), & t \in(0, T] \\[10pt]
        u(x, 0)														 										& =0,          					& x \in[0, L]       \\[10pt]
        \partial_{t}{u}(x, 0)																				& =0,          					& x \in[0, L]       \\[10pt]
        u\left(0,t\right)																					& =S_{l}\left(t\right),         & t>0       \\[10pt]
        \partial_{t}u\left(L, t\right) +c\left(L\right)\partial_{x}u\left(L,t\right) 						& = 0                			& t>0
    \end{array}
    \label{eq:low_contrast}
\end{equation}

The velocity field approximation was performed using $\beta$-splines to describe the behavior of the transverse wave speed squared within the region of interest:

\begin{equation}
    c^2\left(x\right)=\sum_{j=0}^{n-1}\theta_j B_{j,k,\nu}\left(x\right)
    \label{eq:beta_splines}
\end{equation}

Here, $\boldsymbol{\theta}$ represents the vector of coefficients in the $\beta$-splines model, and $B_{j,k,\nu}$ are the basis functions of degree $k$ at knots $\nu$, given by the following recurrence relation:

\begin{equation}
    B_{i,0}(x) = 
    \begin{cases}
    1, & \text{if } \nu_i \leq x < \nu_{i+1}, \\[5pt]
    0, & \text{otherwise},
    \end{cases}
    \label{eq:beta_splines_b0}
\end{equation} 
\begin{equation}
    B_{i,k}(x) = \frac{x - \nu_i}{\nu_{i+k} - \nu_i} B_{i,k-1}(x) + \frac{\nu_{i+k+1} - x}{\nu_{i+k+1} - \nu_{i+1}} B_{i+1,k-1}(x)
    \label{eq:beta_splines_bi}    
\end{equation}

From equations \eqref{eq:beta_splines}, \eqref{eq:beta_splines_b0}, and \eqref{eq:beta_splines_bi}, we derive the relation \( n \geq k+1 \); otherwise, it is not possible to construct the recurrence model. This relation indicates that, given \( n \) coefficients, it is possible to build $\beta$-splines of degrees ranging from \( k=0 \) to \( k=n-1 \). More information about $\beta$-splines can be found in \citep{de1978practical,lyche2008spline}.
We set the parameter space dimension as \( n=15 \) and constructed the 15 associated $\beta$-splines models. Each of these models was tested for the same values of \( \omega \) as in the previous case.
\subsubsection{Bayesian Model}
The Bayesian model implemented in this case is similar to the model proposed in \eqref{eq:modelo_bayesiano_modelo_surrogate}. The solution \( u \) to the forward model is provided by the system in \eqref{eq:low_contrast}, and the observation vector is given by \( \boldsymbol{y}(t) = \left(y_1(t), \dots, y_N(t)\right) \). As in the previous model, \( N \) represents the number of observations within the region of interest.

With this in mind, we have:
\begin{equation}
    \log \pi(\theta \mid \mathbf{y}) \approx -\frac{1}{2\sigma^{2}} \Vert \mathbf{y}(t) - u\left(\mathbb{X}, t \right) \Vert_{L\left[0, T\right]^{n}}^{2} + \left( \theta - \theta_{0} \right)^\intercal \mathbf{M}^{-1} \left( \theta - \theta_{0} \right)
\end{equation}

\begin{equation}
    \begin{aligned}
        -\frac{1}{2\sigma^{2}} \Vert \mathbf{y}(t) - u\left(\mathbb{X}, t \right) \Vert_{L\left[0, T\right]^{n}}^{2} &= \int_{0}^{T} -\frac{1}{2\sigma^{2}} \left( \mathbf{y}(t) - u\left(\mathbb{X}, t \right) \right)^{2} \, dt
    \end{aligned}
\end{equation}

where
\begin{equation}
    u\left(\mathbb{X}, t \right) = \left(u\left(x_1, t \right), \dots, u\left(x_N, t \right) \right)
\end{equation}

\begin{figure}[H]
    \centering
    \includegraphics[width=0.8\textwidth]{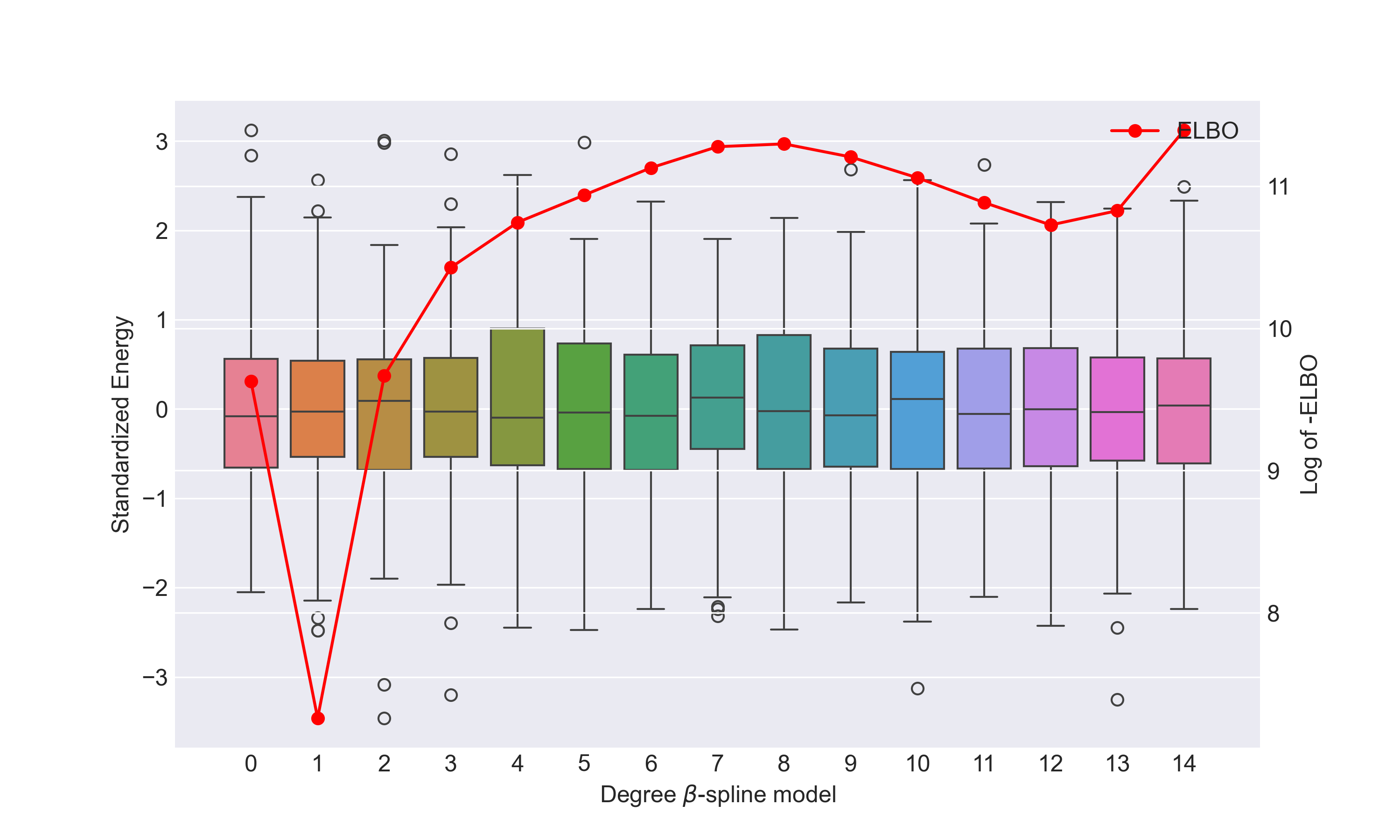}
    \caption{Comparison of energy and log ELBO variation across $\beta$-spline models for different degrees of freedom and \( \omega = 0.0 \). The red dotted line indicates the variation of the log ELBO for each model.}
    \label{fig:w00}
\end{figure}

\begin{figure}[H]
    \centering
    \includegraphics[width=0.8\textwidth]{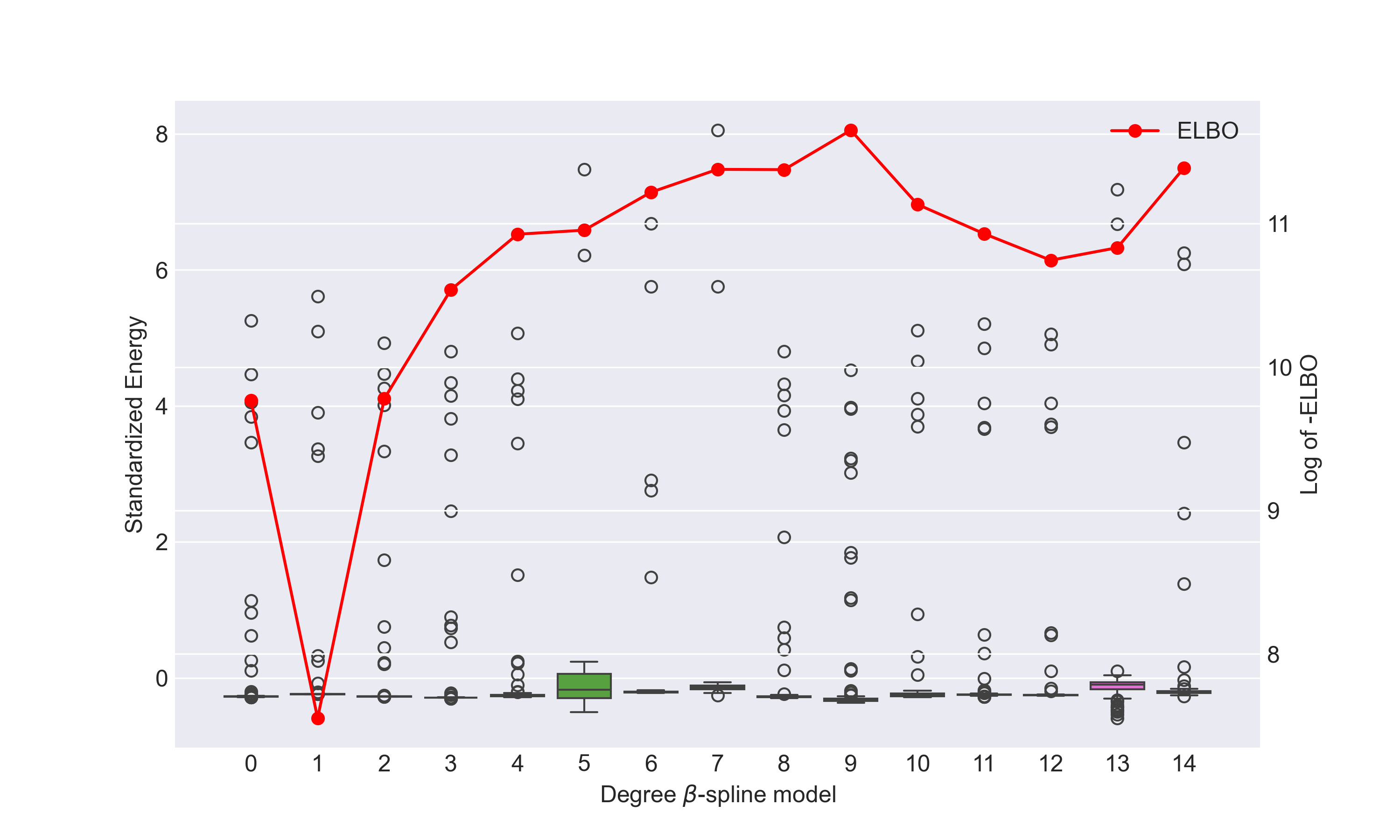}
    \caption{Comparison of energy and log ELBO variation across $\beta$-spline models for different degrees of freedom and \( \omega = 0.5 \). The red dotted line indicates the variation of the log ELBO for each model.}
    \label{fig:w05}
\end{figure}

\begin{figure}[H]
    \centering
    \includegraphics[width=0.8\textwidth]{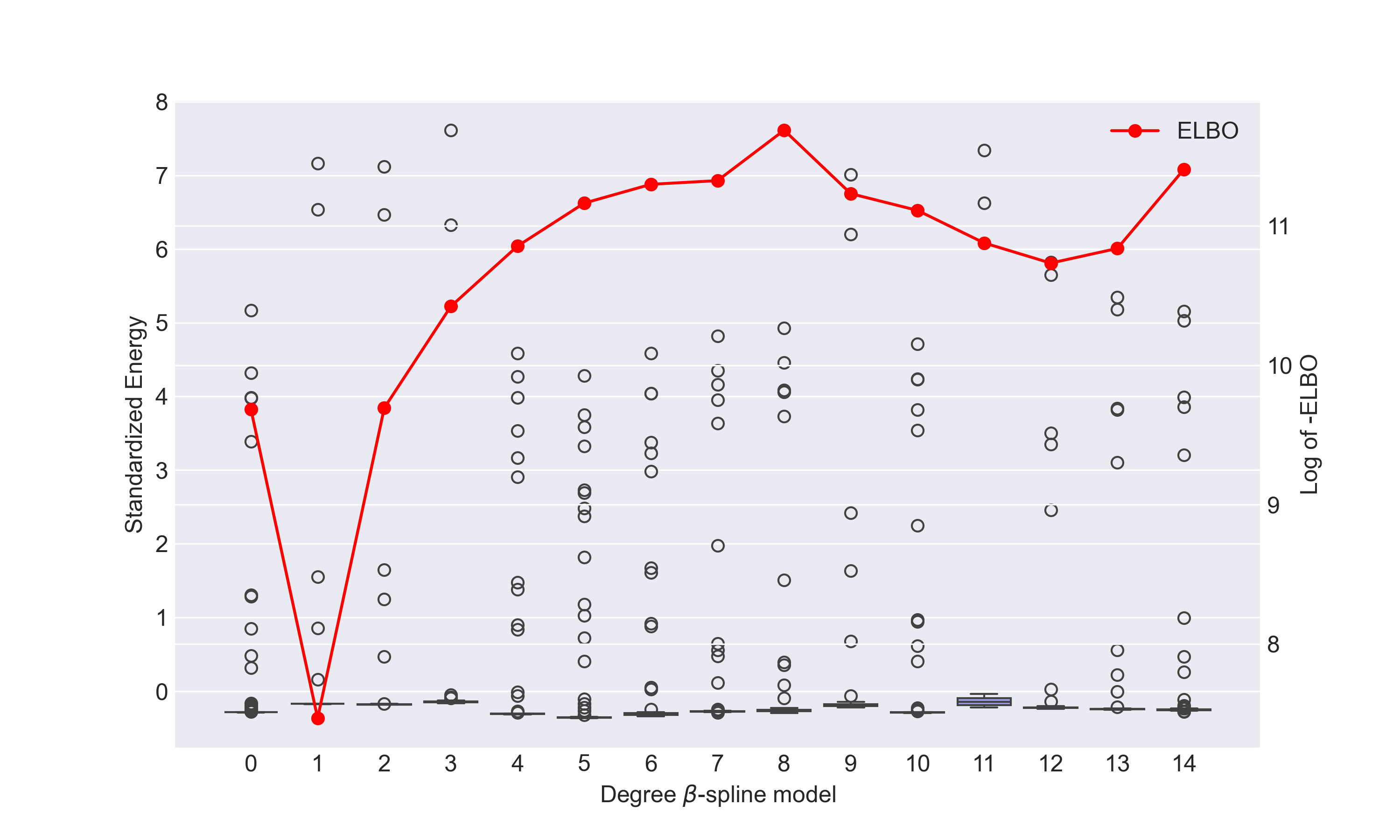}
    \caption{Comparison of energy and log ELBO variation across $\beta$-spline models for different degrees of freedom and \( \omega = 1 .0 \). The red dotted line indicates the variation of the log ELBO for each model.}
    \label{fig:w10}
\end{figure}

\begin{figure}[htbp]
    \centering
    \begin{subfigure}[b]{0.48\textwidth}
        \centering
        \includegraphics[width=\linewidth]{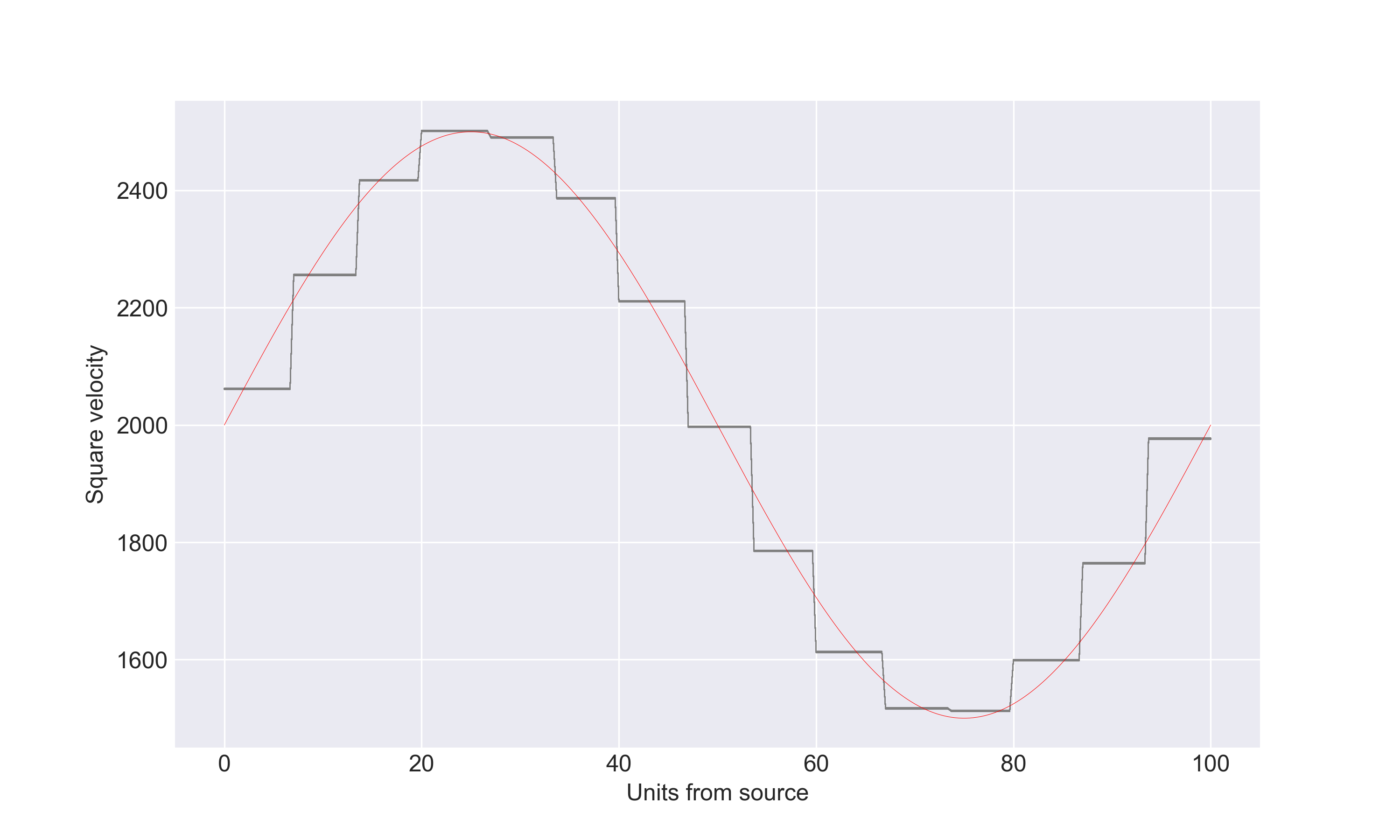}
        \caption{$k=0$}
    \end{subfigure}
    \hfill
    \begin{subfigure}[b]{0.48\textwidth}
        \centering
        \includegraphics[width=\linewidth]{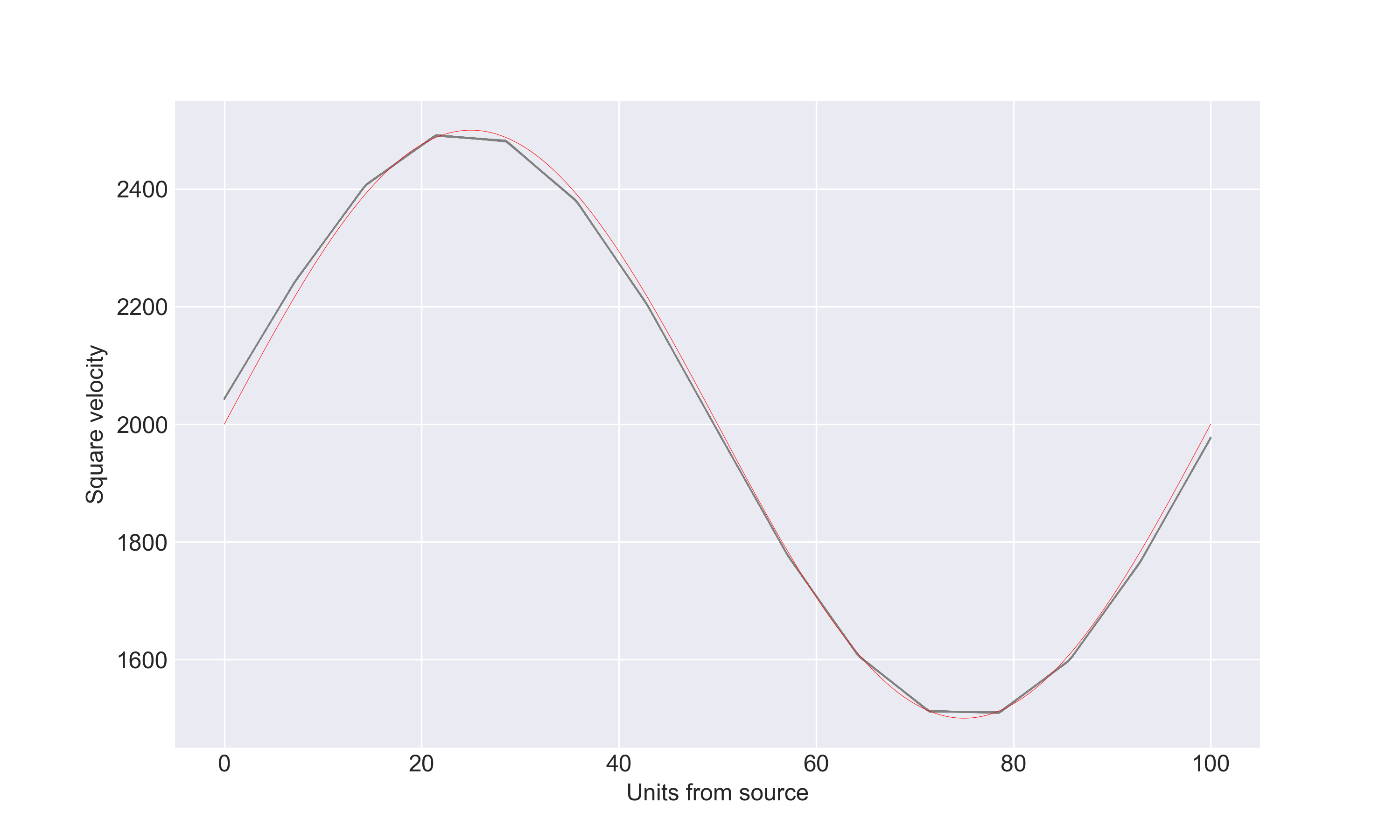}
        \caption{$k=1$}
    \end{subfigure}

    \vspace{0.3cm} 

    \begin{subfigure}[b]{0.48\textwidth}
        \centering
        \includegraphics[width=\linewidth]{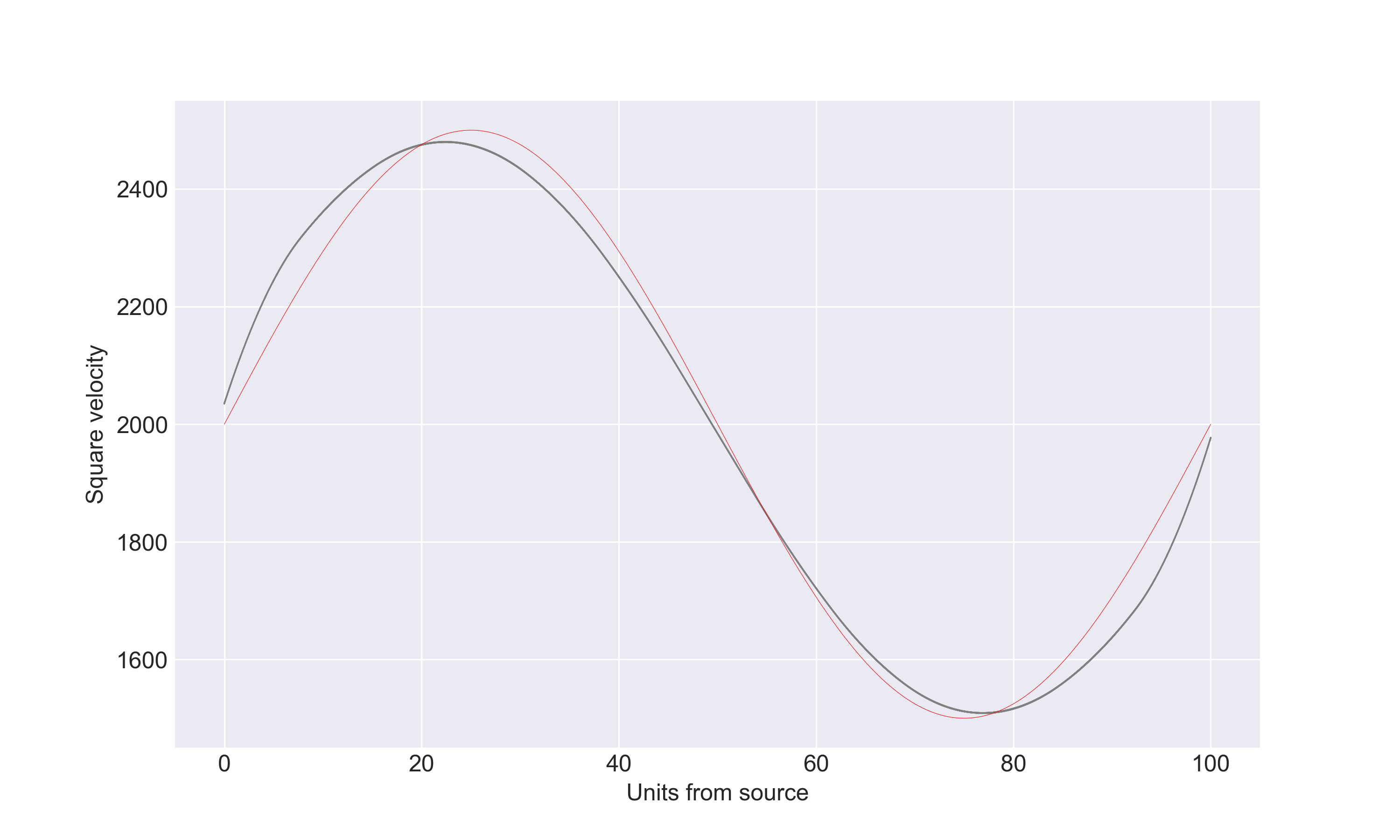}
        \caption{$k=2$}
    \end{subfigure}
    \hfill
    \begin{subfigure}[b]{0.48\textwidth}
        \centering
        \includegraphics[width=\linewidth]{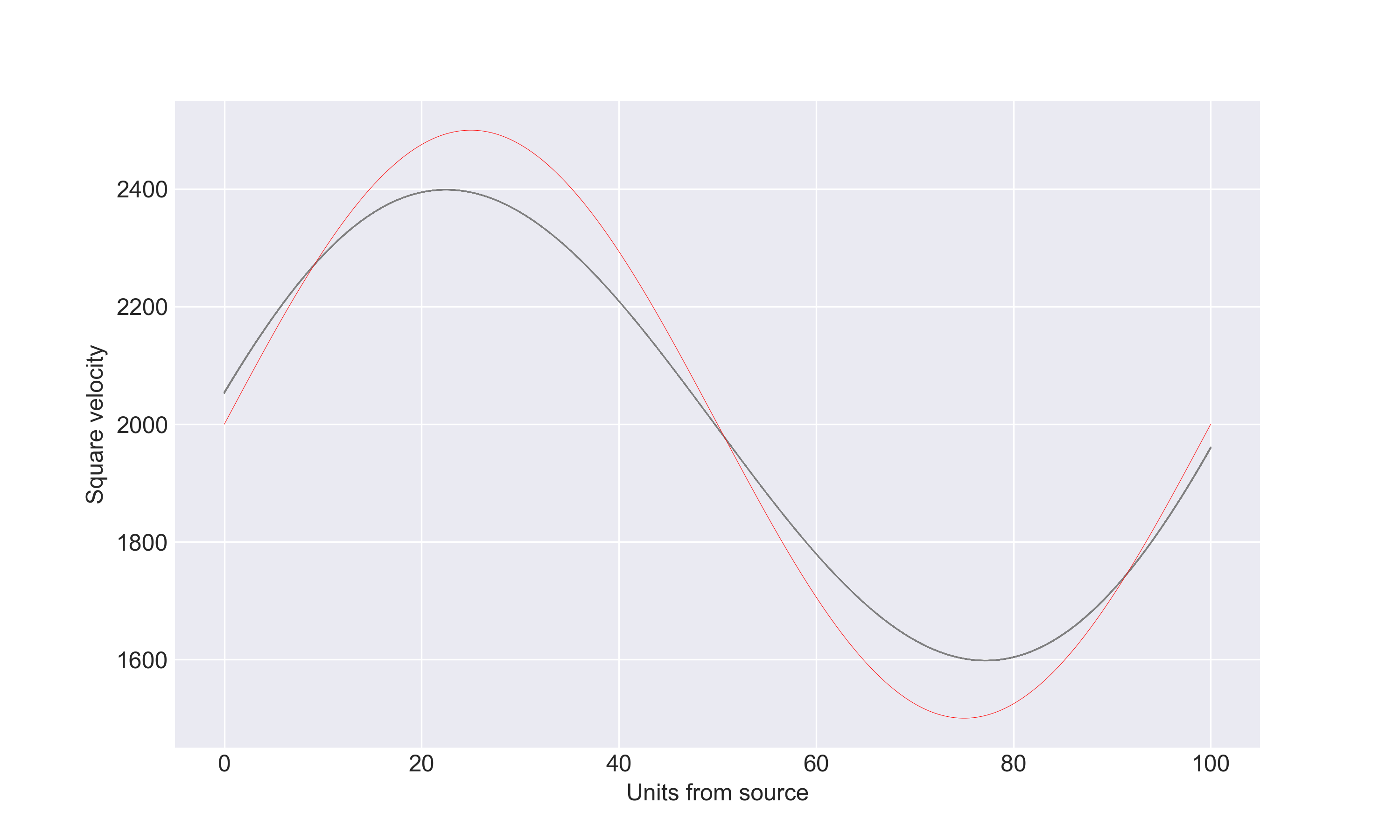}
        \caption{$k=14$}
    \end{subfigure}
    
    \caption{Comparison of velocity field approximations for different values of $k$ with $\omega=0.0$. The images show cases for $k=0$, $k=1$, $k=2$, and $k=14$.}
    \label{fig:velocity_comparison_w00}
\end{figure}

\begin{figure}[htbp]
    \centering
    \begin{subfigure}[b]{0.48\textwidth}
        \centering
        \includegraphics[width=\linewidth]{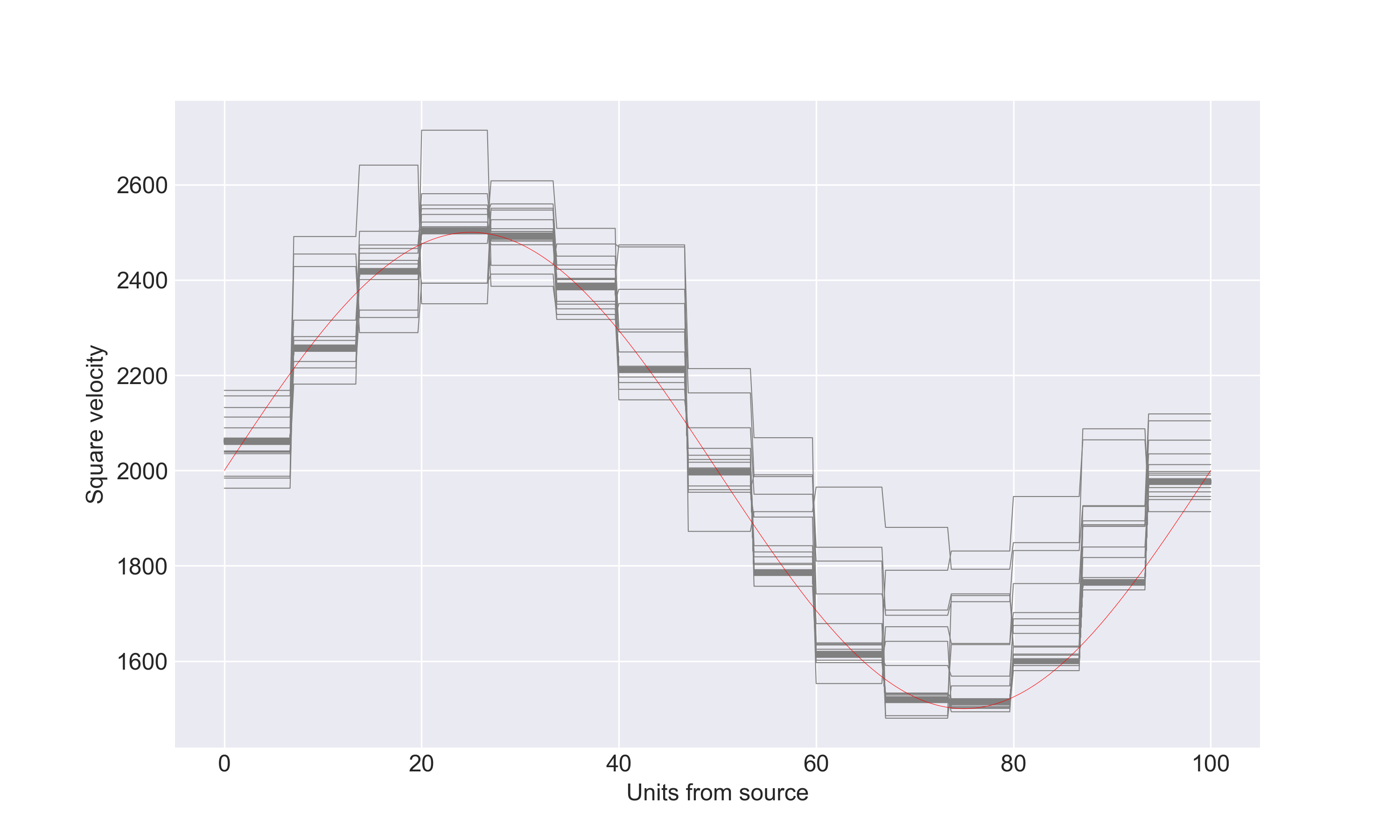}
        \caption{$k=0$}
    \end{subfigure}
    \hfill
    \begin{subfigure}[b]{0.48\textwidth}
        \centering
        \includegraphics[width=\linewidth]{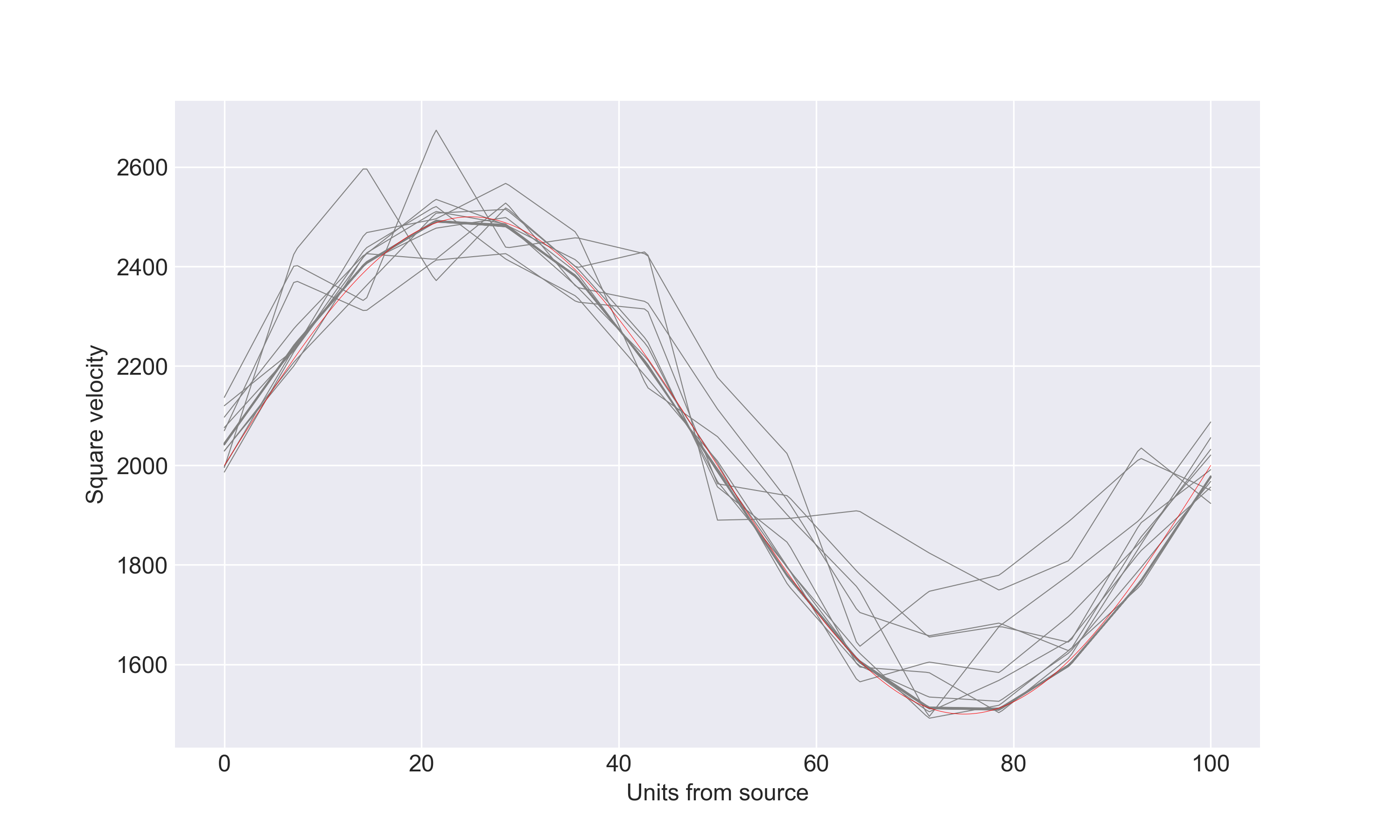}
        \caption{$k=1$}
    \end{subfigure}

    \vspace{0.3cm} 

    \begin{subfigure}[b]{0.48\textwidth}
        \centering
        \includegraphics[width=\linewidth]{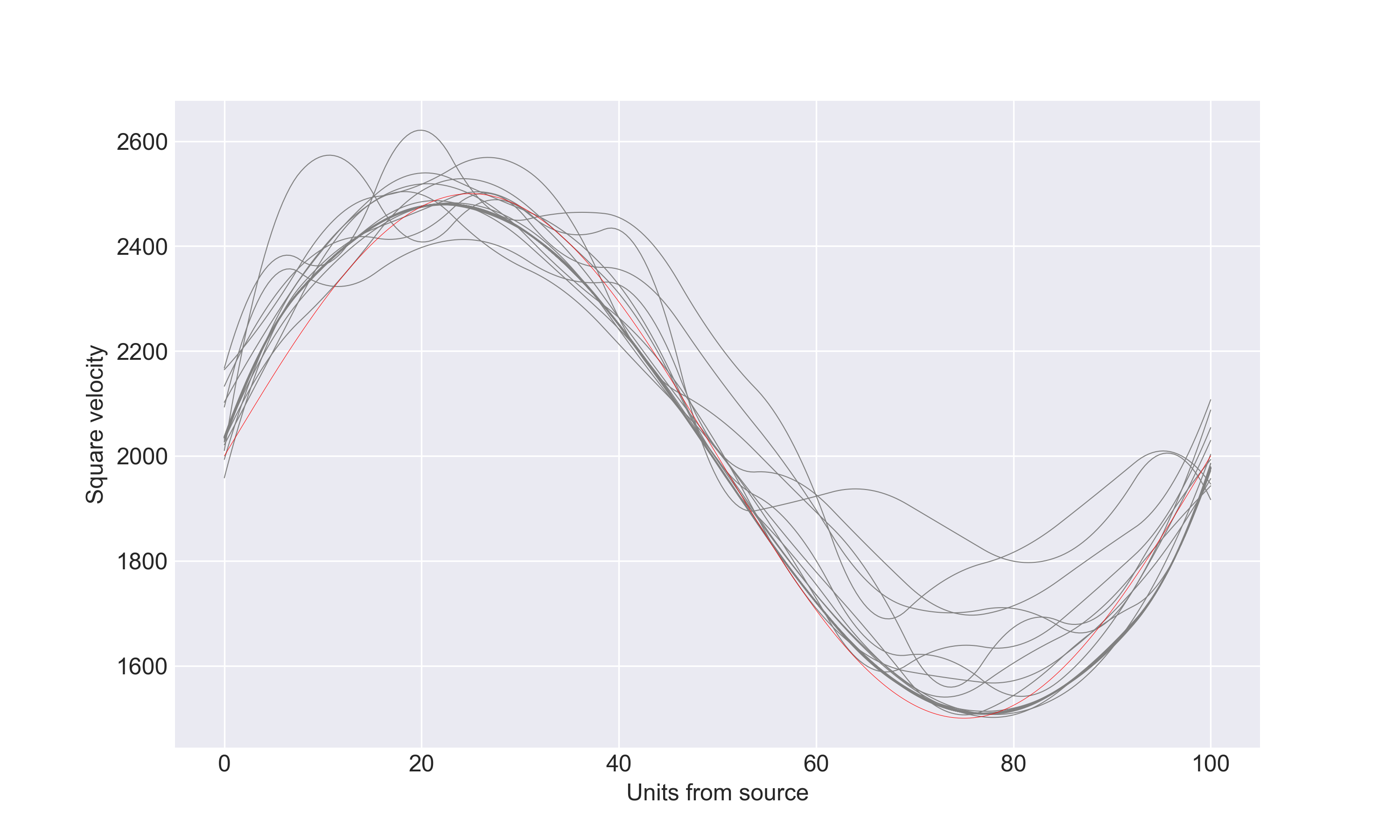}
        \caption{$k=2$}
    \end{subfigure}
    \hfill
    \begin{subfigure}[b]{0.48\textwidth}
        \centering
        \includegraphics[width=\linewidth]{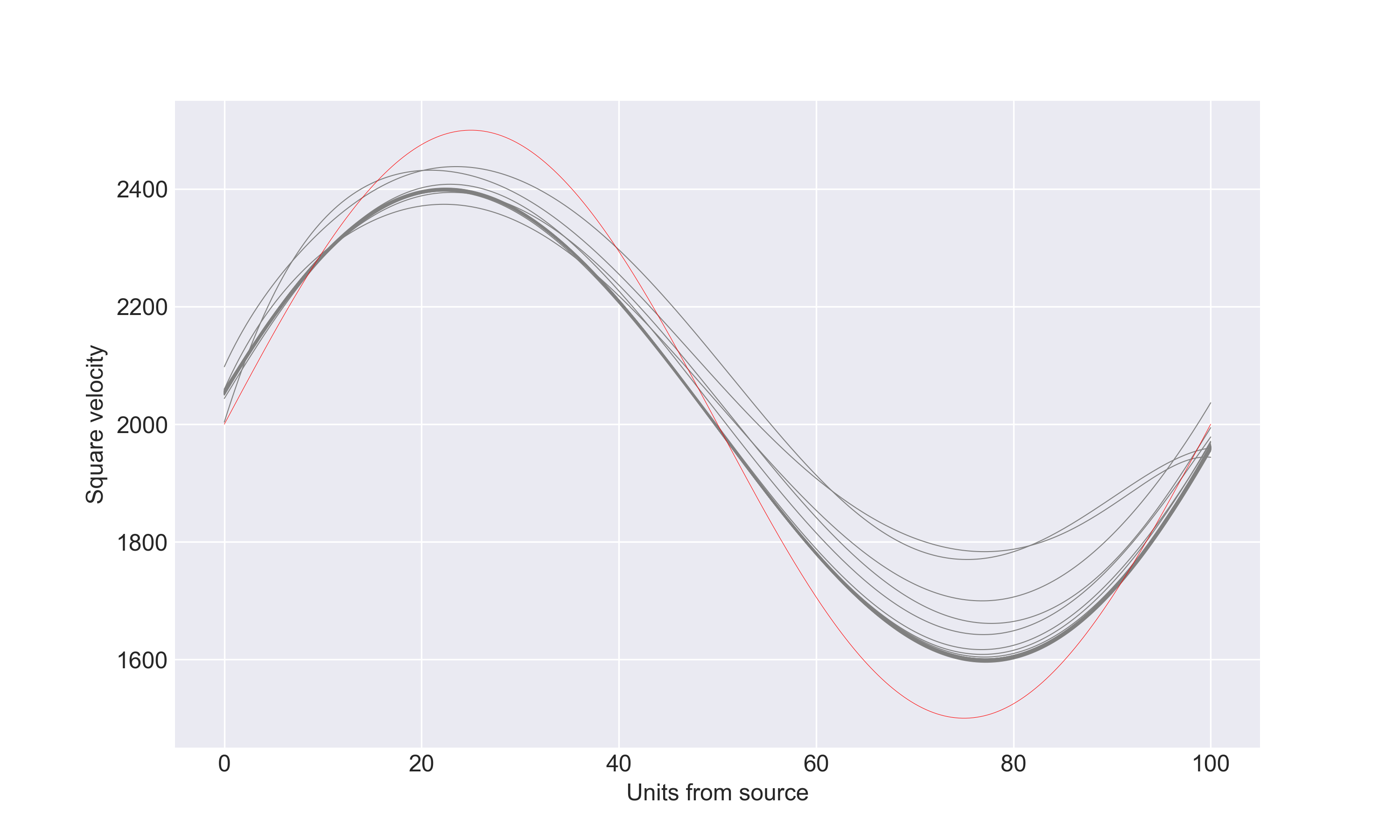}
        \caption{$k=14$}
    \end{subfigure}
    
    \caption{Comparison of velocity field approximations for different values of $k$ with $\omega=0.5$. The images show cases for $k=0$, $k=1$, $k=2$, and $k=14$.}
    \label{fig:velocity_comparison_w05}
\end{figure}

\begin{figure}[htbp]
    \centering
    \begin{subfigure}[b]{0.48\textwidth}
        \centering
        \includegraphics[width=\linewidth]{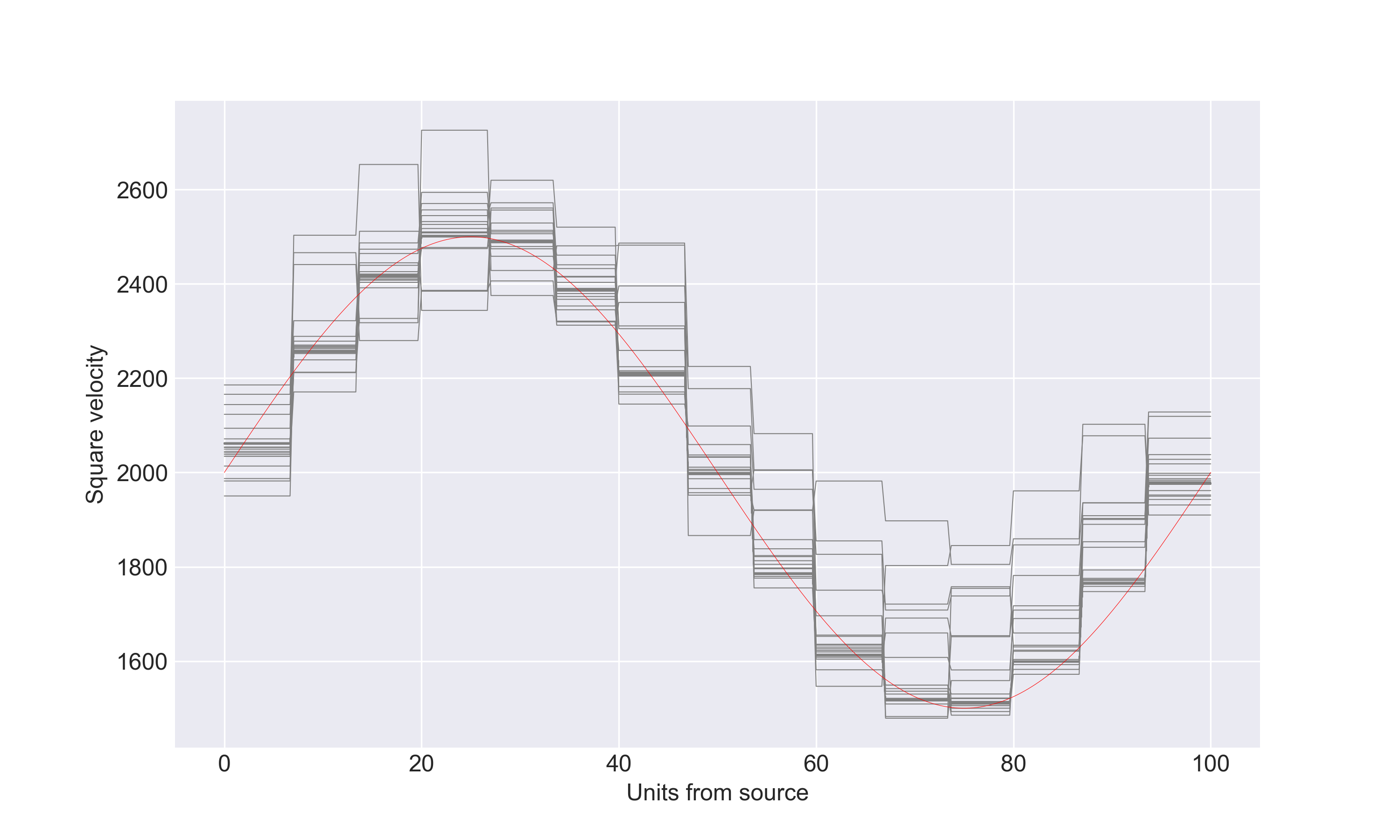}
        \caption{$k=0$}
    \end{subfigure}
    \hfill
    \begin{subfigure}[b]{0.48\textwidth}
        \centering
        \includegraphics[width=\linewidth]{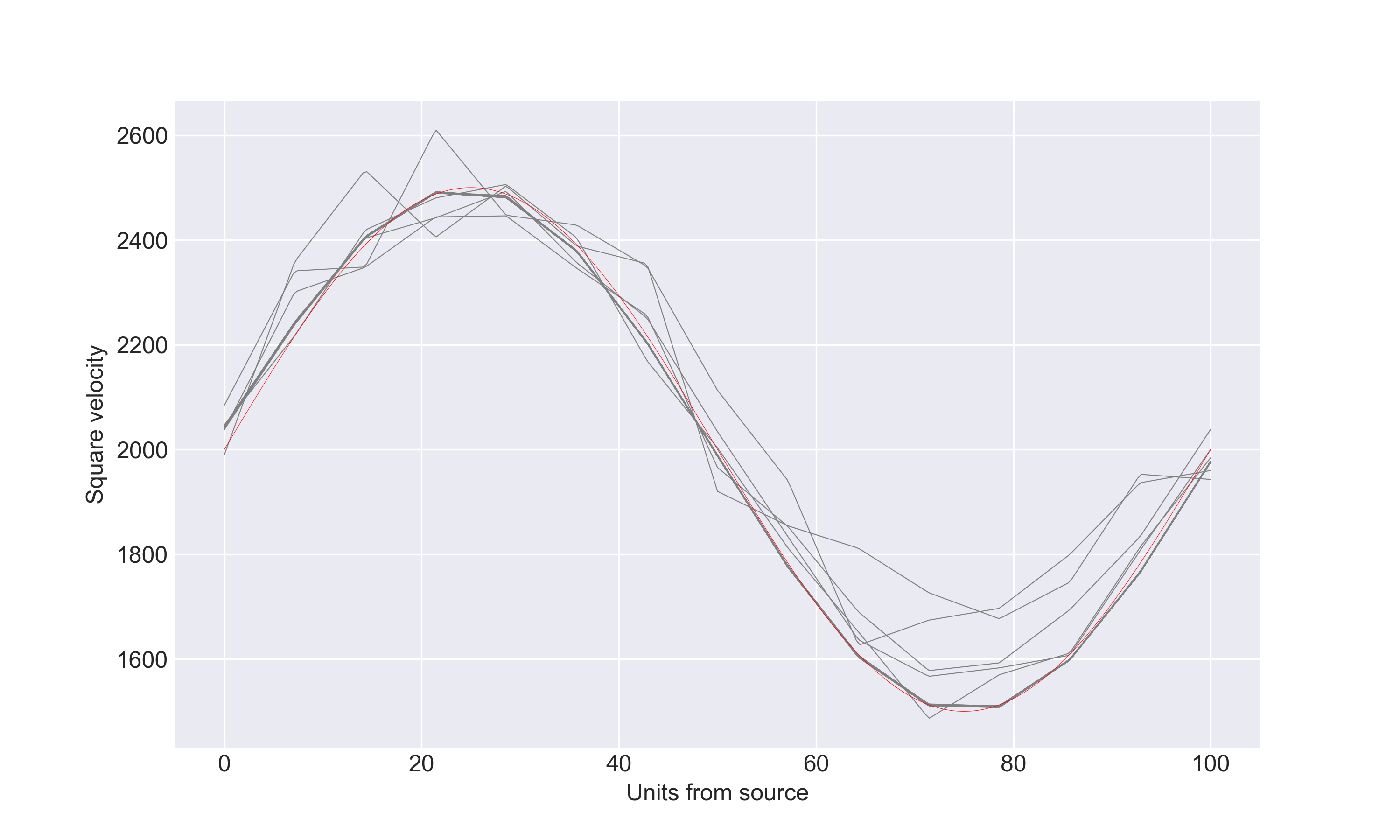}
        \caption{$k=1$}
    \end{subfigure}

    \vspace{0.3cm} 

    \begin{subfigure}[b]{0.48\textwidth}
        \centering
        \includegraphics[width=\linewidth]{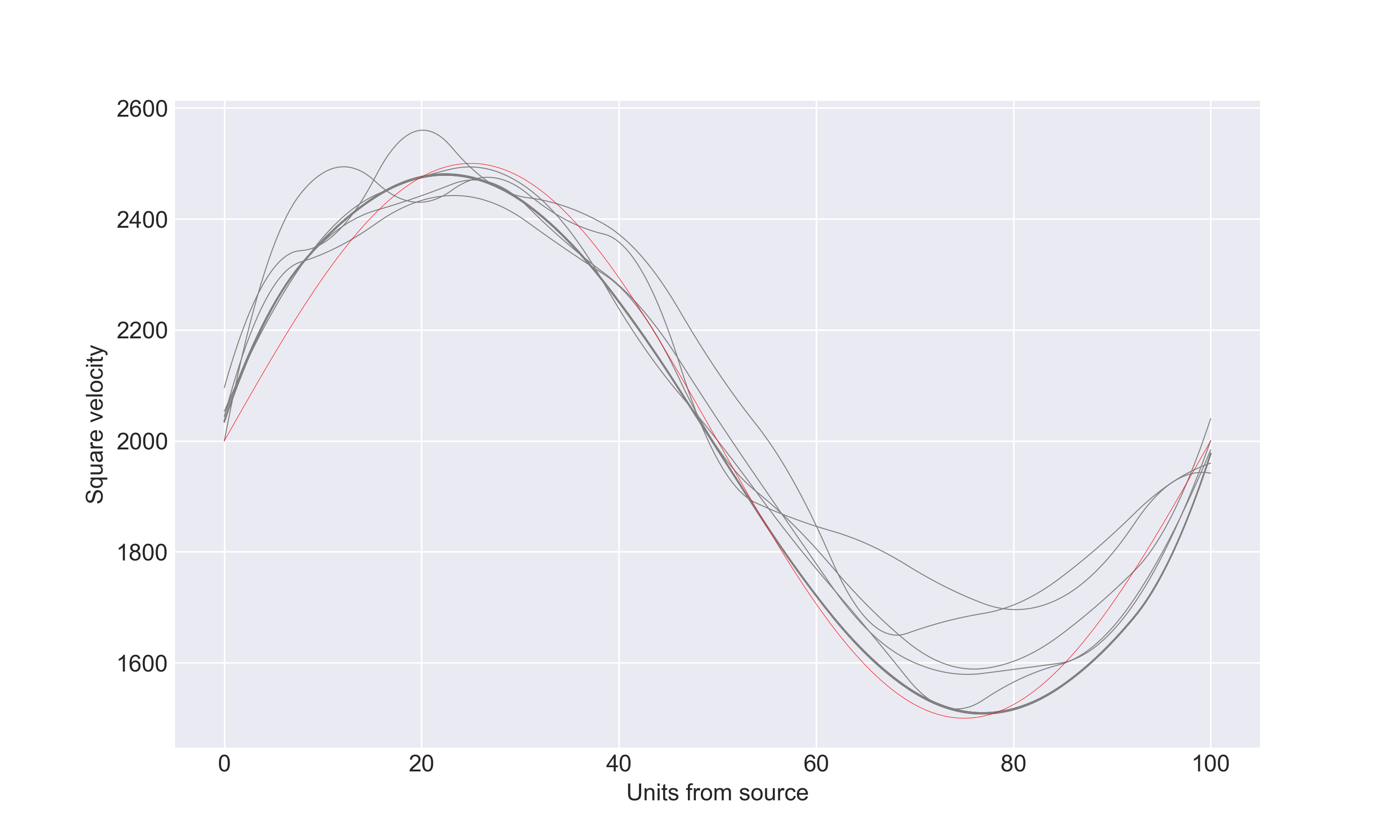}
        \caption{$k=2$}
    \end{subfigure}
    \hfill
    \begin{subfigure}[b]{0.48\textwidth}
        \centering
        \includegraphics[width=\linewidth]{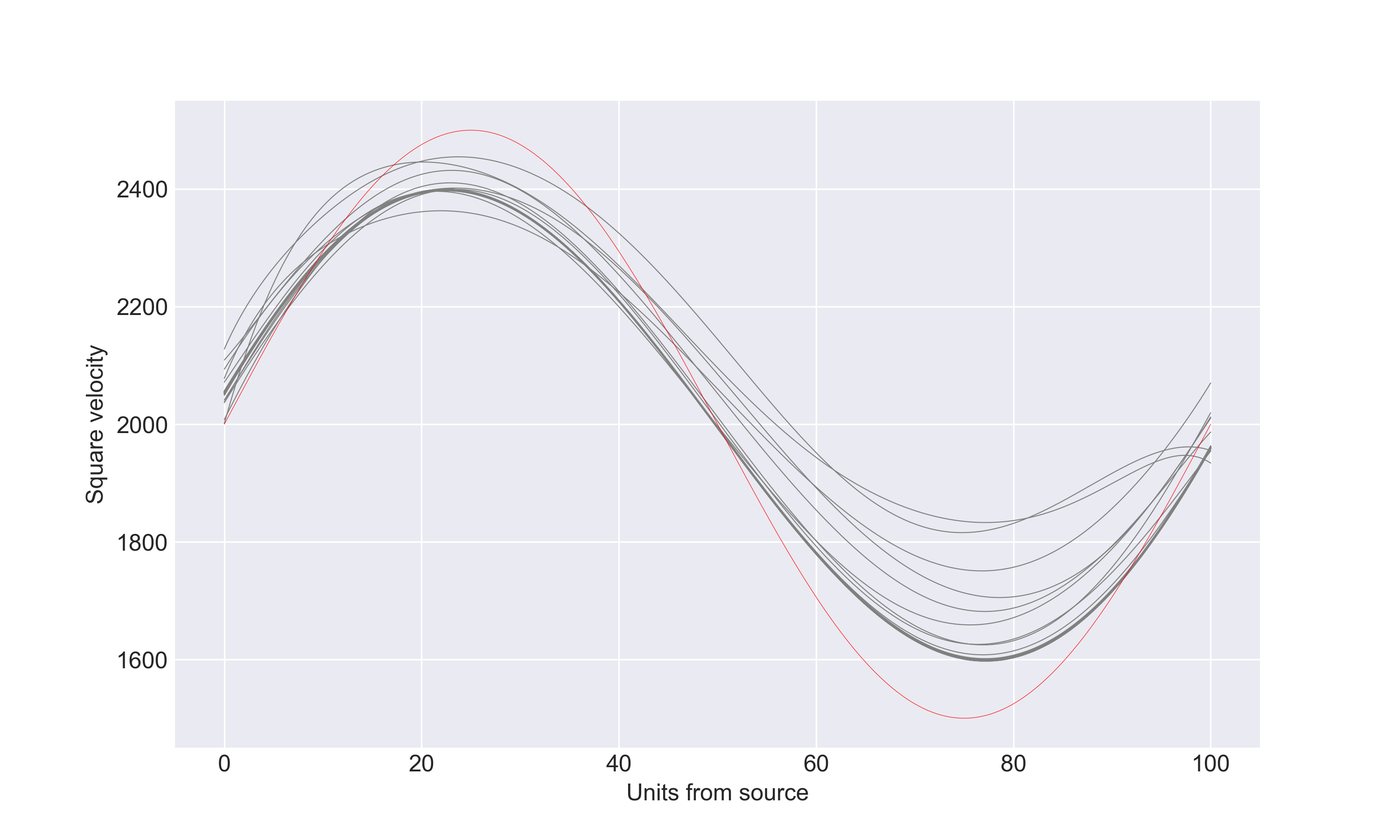}
        \caption{$k=14$}
    \end{subfigure}
    
    \caption{Comparison of velocity field approximations for different values of $k$ with $\omega=1.0$. The images show cases for $k=0$, $k=1$, $k=2$, and $k=14$.}
    \label{fig:velocity_comparison_w10}
\end{figure}

The simulation results are shown in Figures \ref{fig:w00}, \ref{fig:w05}, and \ref{fig:w10}. These figures contain a selection of the most significant results from the simulations. See Figures \ref{fig:velocity_comparison_w00},\ref{fig:velocity_comparison_w05} and \ref{fig:velocity_comparison_w10} for the comparsion of the posterior predective among the differents values of $\omega$ and degree of freedom for the $\beta$-splines models.

\clearpage

\section{Discussion}\label{sec:discusion}

\subsection{Problem and Importance}

This work focuses on wave propagation in high- and low-contrast stratified media, with applications in Geophysics, where accurate velocity modeling is crucial for subsurface structure interpretation and material characterization. Accurate wave modeling in heterogeneous media is key for seismic exploration, resource assessment, and geological monitoring. Precise velocity fields improve seismic interpretation, reducing costs and enhancing exploration efficiency.

Our model addresses the 1D case of two wave prospection models, but can be extended to 2D and 3D, supporting broader applications in geophysical exploration.


\subsection{Contributions and Problems Solved}

This research yielded several key contributions. First, we employed the five-point operator for approximating first- and second-order derivatives. While this operator has previously been used in geophysical models \citep{igel2017computational}, our approach specifically focuses on sinusoidal and exponential decay functions. This allowed us to determine the number of nodes per wavelength required for the numerical model to closely mimic the physical model. The results of this approximation are shown in Figures \ref{fig:relative_error} and \ref{fig:derivative_approximation}.

Another application of the five-point operator was in handling absorbing boundary conditions. The differential equations for accurate absorbing boundary approximations, as derived in \citep{buckman2012onesided}, include first- and second-order differential equations. In this study, we implemented their first-order approximation. Using backward finite differences and centered finite differences for these conditions can introduce numerical errors. However, applying the five-point operator with the equations in \eqref{eq:ecuaciones_5p} and stencils $\phi_{0}$ and $\phi_{N}$ minimize these errors, leading to more precise numerical results. Figure \ref{fig:comparacion_diferencias_finitas} compares the boundary solutions obtained, demonstrating that the five-point operator achieves numerical stability more quickly, with fewer degrees of freedom per wavelength.

In addition, we provide a theoretical bound for the error in the numerical approximation of the derivatives until 4th degree using the five points operator. This result can be extended to higher derivative order if needed solving a linear system of equation. 

Through these contributions, our numerical model closely represents the differential models proposed in \eqref{eq:DF1DWE_usr}, \eqref{eq:DF1DWE_usl}, and \eqref{eq:low_contrast}.

Another contribution of this work is in the design of the loss function \eqref{eq:loss_function}. An essential aspect of applying gradient descent methods is defining an effective stopping condition, as an incorrect choice can lead to model overfitting. Additionally, multimodality is a critical factor. The proposed loss function addresses these challenges by ensuring that each sample converges to its nearest local mode at iteration \( i \). This is achieved by considering the maximum gradient magnitude across all samples at state \( i \), providing a measure of the proximity of samples to their local modes. Furthermore, this function incorporates the ELBO, which provides crucial information about the distribution \citep{blei2017variational,ronning2023elbo}. In \citep{blei2017variational}, a parametric family of distributions is used to approximate the posterior distribution; however, this approach faces challenges in handling multimodality in more complex distributions. We use KDE to provide this approximation at each iteration step, aiming to create a Gaussian approximation of the distribution around each local mean at every state.

Using this approach, we developed the G-SVGD algorithm, whose core is based on Algorithm 1, \textit{Bayesian Inference via Variational Gradient Descent}, as described in \citep{liu2016stein}, with three essential modifications. As described in \citep{liu2016stein}, the algorithm requires three key elements to converge to a sample of the posterior distribution. 

First, it needs an initial sample that is sufficiently close to the local modes; if this initial sample is too distant, the algorithm may fail in the transformation process. Second, the step size, or learning rate, must be carefully chosen to ensure a smooth transition to the next step. Third, the stop condition is not specified by the author and must be defined to avoid unnecessary computation or convergence issues.

We address these three elements as follows. For the initial sample, we use the quasi-Monte Carlo Halton algorithm \citep{owen2017randomized}. This algorithm is also applied effectively to optimize observation nodes in linear inverse problems, as demonstrated in \citep{rojo2020optimo}. For the step size, we propose an optimization method based on the loss function, selecting a single, global step size for the entire sample at each state \( i \). A similar approach is used in \citep{futami2019bayesian}, where the author employs the Frank-Wolfe algorithm \citep{frank1956algorithm} to approximate the objective function linearly, assigning an individual step size for each particle in the sample to guide convergence toward the MAP estimate.

Our approach differs in the following ways:
\begin{enumerate}
    \item \textbf{Step Size Determination:} Our step size proposal is informed by the sample's loss function, enabling a global step size for all samples rather than individualized particle-based optimization.
    \item \textbf{Local Distribution Approximation:} We employ KDE to approximate a Gaussian distribution around each sample at every state, providing a probabilistic representation of the distribution around local means—something not explicitly addressed in \citep{futami2019bayesian}.
    \item \textbf{Stopping Condition:} We define the stop condition using the loss function to prevent unnecessary iterations and ensure convergence.
\end{enumerate}

These modifications are encapsulated in the G-SVGD, enhancing the robustness and applicability of our approach in sampling from multi-modal distributions. Additionally, the author in \citep{liu2016stein} utilized the ADAM method \citep{kingma2014adam} to determine the step size at each state. However, this is not the only method available for selecting an adaptive learning rate. In \citep{ruder2016overview}, the author provides an extensive discussion on different methods used in machine learning for this purpose. 

We employ the ADAM method as a comparison approach based on the following criteria:
\begin{enumerate}
    \item \textbf{Effectiveness in Multimodal Approximation:} The author in \citep{liu2016stein} successfully applied ADAM to approximate a multimodal normal distribution, similar to the Gaussian approximation we aim to achieve in our method.
    \item \textbf{Adaptive Learning Rates for Each Parameter:} ADAM computes an adaptive learning rate for each parameter, allowing simultaneous optimization across subspaces of the parameter space.
\end{enumerate}

We applied these ideas to the high-contrast medium. The numerical results for the approximation in the 2-dimensional parameter space are shown in Figure \ref{fig:histograms}. This figure demonstrates how variations in the $\omega$ parameter directly impact the accuracy of our method’s approximation, making $\omega$ a critical parameter to consider in simulations.

The left column of Figure \ref{fig:histograms} presents the simulation results for the different values of $\omega$ selected in this experiment. Our method successfully identifies the local modes, with variations observed in the distribution shape approximation depending on $\omega$. The right column shows results from the reference methods: t-walk \citep{christen2010general} and ADAM Stein Variational Gradient Descent (A-SVGD).

The t-walk is an MCMC-based method that uses random walks and operates independently of initial samples, step size, and stopping conditions. Its robustness makes it a widely used choice for approximating complex multimodal distributions, which makes it an ideal control group in our numerical experiment.

A-SVGD uses a similar approach to our G-SVGD algorithm, with a notable difference in step size selection. This comparison allows us to evaluate two key aspects: the efficiency of our step size proposal based on optimization, and whether our loss function is effectively guiding the method in the desired direction.

The results of these approaches are shown in the right column of Figure \ref{fig:histograms}. The t-walk approach converges to a similar solution as G-SVGD for values of $\omega = \{0.0, 0.5, 1.0\}$. However, t-walk required 10,000 samples to properly capture the objective distribution, while our G-SVGD approach achieved similar accuracy using only 200 samples, leading to a significant improvement in computational cost, given that the model complexities are equivalent.
A-SVGD method localized the local modes, but completly ignore the shape of the distribution. This is  becuase this method doesn't take in account any filter to approximate the distribution itself.

The final part of this experiment involved comparing the results. Since the loss function is sensitive to the values of \( \omega \), results from different values of this parameter are not directly comparable. However, they can be compared with the results provided by the t-walk approximation and A-SVGD by evaluating the final numerical results using the same loss function. Figures \ref{fig:convergence_w00}, \ref{fig:convergence_w05}, and \ref{fig:convergence_w10} show the convergence of the numerical results relative to the reference values obtained with the t-walk approach.

In the cases \(\omega=\{0.0, 0.5\}\), we observe an improvement in the loss function values compared to the t-walk results. This suggests that the loss function serves as an effective measure for approximating the posterior distribution, as minimizing it leads to convergence of the initial sample towards a sample from the posterior distribution.

For the case \(\omega=1.0\), where the ELBO has the full weight in the approximation, the quality of the test distribution (KDE) significantly influences the identification of the shape and location of local modes. Despite this, the iterative results still approach the control values. Additionally, comparing our method with A-SVGD in terms of convergence to the control value shows that G-SVGD achieves a faster convergence rate in all three figures. This indicates that our optimization-based approach, which minimizes the loss function at each step, has a substantial impact on reducing computational cost. Furthermore, the proposed model effectively fulfills its intended role.

The second prospection model focuses on approximating the velocity field in a low-stratified medium. Using \(\beta\)-spline models, we redefine the interpretation of the parameter space. Unlike the previous examples, where parameters directly determined the velocity, this approach uses coefficients of the \(\beta\)-spline model as parameters. This adjustment allows for an increased parameter space dimension, enhancing approximation quality without increasing computational cost. Now, we can modify the degrees of freedom for the basis in the \(\beta\)-spline and arrive at conclusions regarding which of them provides a better approximation. Figure \ref{fig:w00}, \ref{fig:w05} and \ref{fig:w10} show the numerical results of the 45 models analyzed. As before, we cannot compare the results for different values of the \(\omega\) parameter. In this case, we used \(\log\left( -\text{ELBO} \right)\) as a comparison measure with KDE as our testing distribution. This comparison was conducted on the sample after the algorithm converged.

For the case of \(\omega=0.0\) (Figure \ref{fig:w00}), we observe that the solutions for each of the models for \(k\) are well-centered with similar standard deviations across all models. This occurs because our loss function forces each sample to converge to its local mode, leaving limited space for exploration of distant regions. 

The other two cases leave more room for exploration, resulting in higher standard deviations compared to the case \(\omega=0.0\), as shown in Figures \ref{fig:w05} and \ref{fig:w10}. Given these results, we were prompted to address the following question: What degree value \(k\) provides the best results among models with the same \(\omega\) value? Here, \(\log\left( -\text{ELBO} \right)\) plays an essential role in the comparison. Figures \ref{fig:w00}, \ref{fig:w05} and \ref{fig:w10} illustrate this analysis.
The results of this experiment show that the case \(k=1\) provides the best results across different values of \(\omega\), establishing that linear models are sufficient to approximate the velocity field in a low-stratified medium when a \(\beta\)-spline model is used for the description.
Due to computational restrictions, we could not use the t-walk or the A-SVGD algorithm, as both required significantly higher computational costs.

Finally, the main result of this work is the development of the G-SVGD algorithm, a novel variational inference approach built upon the foundations of Algorithm 1, as introduced in \citep{liu2016stein}. While maintaining the core principles of the original algorithm, G-SVGD incorporates several critical modifications that significantly enhance its performance and applicability in high-dimensional and multimodal scenarios:

\begin{itemize}
    \item \textbf{Adaptive Step Size Optimization:} Our algorithm introduces an adaptive step size approach informed by a custom-designed loss function. This ensures smoother transitions and efficient convergence, addressing the limitations in step size selection in \citep{liu2016stein}.
    \item \textbf{Kernel Density Estimation  for Local Gaussian Approximation:} By leveraging KDE, we provide a probabilistic approximation of the local distribution around each mode. This addition enables better exploration of complex multimodal posterior distributions.
    \item \textbf{GPU Acceleration:} G-SVGD is designed to leverage GPU parallelization, significantly enhancing computational efficiency for high-dimensional problems. By distributing kernel evaluations and gradient computations across GPU cores, the algorithm achieves superior performance, enabling scalability and real-time sampling for large datasets.
    \item \textbf{Scalability and Computational Efficiency:} Unlike traditional MCMC-based methods, such as the t-walk, G-SVGD achieves comparable accuracy while significantly reducing the number of samples and computational cost. The inclusion of GPU compatibility further boosts its practical applicability in resource-intensive scenarios.
    \item \textbf{Validation in Low- and High-Contrast Mediums:} Through comprehensive numerical experiments, we demonstrated the robustness of G-SVGD across a range of challenging physical models, including velocity fields in stratified media.
\end{itemize}

These enhancements position G-SVGD as a powerful tool for Bayesian inference, particularly in applications requiring efficient and accurate sampling from complex posterior distributions. The success of the algorithm in addressing multimodality, computational efficiency, and adaptability underscores its potential for broader adoption in scientific and industrial applications.

\subsection{Limitations and Future Work}

While the G-SVGD algorithm presents significant advancements in variational inference, several limitations and opportunities for future research remain:

\subsection{Limitations}
\begin{itemize}
    \item \textbf{Computational Dependence on GPU Resources:}
    The algorithm's efficiency heavily relies on access to GPU acceleration for tasks such as kernel density estimation (KDE) and gradient computations. Systems lacking high-performance GPUs may face challenges in scalability and runtime efficiency.

    \item \textbf{Sensitivity to Loss Function Parameters:}
    The performance of G-SVGD is influenced by the choice of parameters in the loss function, particularly \(\omega\). While we demonstrated the impact of \(\omega\) on convergence, selecting an optimal value for complex distributions remains an open problem.

    \item \textbf{Dependence on KDE Accuracy:}
    The approximation quality depends on the KDE used to describe the posterior distribution locally. In highly multimodal distributions, KDE smoothing can introduce bias, impacting the representation of local modes.

    \item \textbf{Scalability to Extremely High-Dimensional Spaces:}
    Although G-SVGD outperforms traditional MCMC methods in terms of computational cost, its application in extremely high-dimensional spaces may still require further optimization and exploration of more efficient kernel methods.
\end{itemize}

\subsection{Future Work}
\begin{itemize}
    \item \textbf{Adaptive Parameter Tuning:}
    Investigate dynamic methods for selecting loss function parameters such as \(\omega\) during the training process. This could enhance robustness across a broader range of distributions.

    \item \textbf{Integration with Alternative Kernel Approaches:}
    Explore the use of advanced kernels, such as deep kernels or spectral kernels, to improve KDE accuracy and reduce bias in multimodal settings.

    \item \textbf{Application to High-Dimensional Inference Problems:}
    Extend the algorithm to address challenges in extremely high-dimensional parameter spaces, including the use of dimension reduction techniques and scalable kernel implementations.

    \item \textbf{Hybrid Algorithms:}
    Combine G-SVGD with other inference methods, such as Hamiltonian Monte Carlo or variational autoencoders, to leverage complementary strengths for more efficient posterior approximation.

    \item \textbf{Benchmarking on Real-World Datasets:}
    Apply G-SVGD to large-scale, real-world problems, such as geophysical inversion, Bayesian neural networks, or complex image processing tasks, to evaluate its practical utility and refine its implementation.

    \item \textbf{Energy-Efficient Computation:}
    Optimize the algorithm for deployment on energy-efficient hardware, such as Tensor Processing Units (TPUs) or low-power GPUs, to extend its accessibility and reduce its environmental impact.
\end{itemize}

By addressing these limitations and exploring the proposed directions, the G-SVGD algorithm has the potential to become a more robust and versatile tool for variational inference and Bayesian computation across a wide range of applications.

\begin{description}
    \item[Acknowledgements:] The author JLVS acknowledges to the National Council of Science and Technology of Mexico (CONACyT) under the Ph.D Scholarship program at the Center for Research in Mathematics (CIMAT) 2020-2025.
\end{description}

\begin{description}
    \item[Declaration of competing interest:] The authors declare that they have no known competing financial interests or personal relationships that could have influenced the work reported in this paper.
\end{description}

\begin{description}
    \item[Credit author statement:] \textbf{José L. Varona-Santana} Conceptualization, Formal analysis, Investigation, Methodology, Project administration, Resources, Software, Supervision, Validation, Visualization, Roles/Writing  original draft, Writing  review \& editing; \textbf{Marcos A. Capistrán} Conceptualization, Methodology, Project administration, Resources, Supervision, Writing  review \& editing, Visualization, Roles/Writing  original draft
\end{description}

\begin{description}
    \item[Declaration of generative AI and AI-assisted technologies in the writing process:] During the preparation of this work, the author(s) used ChatGPT and Grammarly in order to improve the clarity, grammar, and structure of the manuscript. After using this tool/services, the author(s) reviewed and edited the content as needed and take(s) full responsibility for the content of the publication.
\end{description}

\clearpage

\bibliographystyle{elsarticle-harv}
\bibliography{references}

\end{document}